\documentclass[twocolumn,superscriptaddress,prb,linenumbers]{revtex4-1}

\usepackage{graphicx,xcolor,amssymb,amsmath,bm,bbm,enumerate}

\newcommand{\ket}[1]{\vert#1\rangle}
\newcommand{\bra}[1]{\langle#1\vert}

\newcommand{\MPS}[1]{| {\bm\mu}[#1]\rangle}
\newcommand{\mc}[1]{\mathcal{#1}}
\newcommand{\mf}[1]{\mathfrak{#1}}
\newcommand{\tr}{\mathrm{tr}}
\newcommand{\dd}{\mathrm{d}}

\newcommand{\sss}[1]{\subsubsection{#1}}
\newcommand{\flip}{\mathbb{F}}

\newenvironment{packed_enum}{
\vspace{-6pt}\begin{enumerate}[i)]
  \setlength{\itemsep}{2pt}
  \setlength{\itemindent}{0pt}
  \setlength{\parskip}{0pt}
  }{\end{enumerate}
  \vspace{-4pt}}

\begin{document}
\nolinenumbers

\title{Classifying quantum phases using MPS and PEPS}
\author{Norbert Schuch}
\address{
Institute for Quantum Information, 
California Institute of Technology, 
MC 305-16,
Pasadena CA 91125, U.S.A.}
\author{David P\'erez-Garc\'ia}
\address{Dpto.\ Analisis Matematico and IMI,
Universidad Complutense de Madrid, 28040 Madrid, Spain}
\author{Ignacio Cirac}
\address{Max-Planck-Institut f\"ur Quantenoptik, Hans-Kopfermann-Str.\ 1, 
D-85748 Garching, Germany}

\begin{abstract}
We give a classification of gapped quantum phases of one-dimensional
systems in the framework of Matrix Product States (MPS) and their
associated parent Hamiltonians, for systems with unique as well as
degenerate ground states, and both in the absence and presence of
symmetries.  We find that without symmetries, all systems are in the same
phase, up to accidental ground state degeneracies. If symmetries are
imposed, phases without symmetry breaking (i.e., with unique ground
states) are classified by the cohomology classes of the symmetry group,
this is, the equivalence classes of its projective representations, a
result first derived in [X.~Chen, Z.-C.~Gu, and X.-G.~Wen, Phys. Rev. B
\textbf{83}, 035107 (2011); arXiv:1008.3745].  For phases with symmetry
breaking (i.e., degenerate ground states), we find that the symmetry
consists of two parts, one of which acts by permuting the ground states,
while the other acts on individual ground states, and phases are labelled
by both the permutation action of the former and the cohomology class of
the latter.  Using Projected Entangled Pair States (PEPS), we subsequently
extend our framework to the classification of two-dimensional phases in
the neighborhood of a number of important cases, in particular systems
with unique ground states, degenerate ground states with a local order
parameter, and topological order.  We also show that in two dimensions,
imposing symmetries does not constrain the phase diagram in the same way
it does in one dimension.  As a central tool, we introduce the
\emph{isometric form}, a normal form for MPS and PEPS which is a
renormalization fixed point.  Transforming a state to its isometric form
does not change the phase, and thus, we can focus on to the classification
of isometric forms.  
\end{abstract}

\maketitle


\section{Introduction}

\subsection{Background}

Understanding the phase diagram of correlated quantum many-body systems,
that is, the different types of order such systems can exhibit, is
one of the most important and challenging tasks on the way to a
comprehensive theory of quantum many-body systems. Compared to classical
statistical models,
quantum systems exhibit a much more complex behavior, such as phases
with topological order as in the fractional quantum Hall effect, which
cannot be described using Landau's paradigm of local symmetry breaking.

In the last years, tensor network based ansatzes, such as Matrix
Product States~\cite{fannes:FCS} (MPS) and Projected Entangled Pair
States~\cite{verstraete:2d-dmrg} (PEPS), have proven increasingly
successful in describing ground states of quantum many-body systems. In
particular, it has been shown that MPS and PEPS can approximate ground
states of gapped quantum systems
efficiently:~\cite{verstraete:faithfully,hastings:arealaw,hastings:mps-entropy-ent}
this is, not only the ground state of systems with local order, but also,
for instance, the ground states of topological insulators are well
represented by those states. Since MPS and PEPS provide a
characterization of quantum many-body states from a local description,
they are promising candidates for a generalization of Landau's
theory. Moreover, they can be used to construct 
exactly solvable models, as every MPS and PEPS appears as the exact
ground state of an associated parent Hamiltonian.~\cite{fannes:FCS,%
nachtergaele:degen-mps,perez-garcia:mps-reps,perez-garcia:parent-ham-2d}

In this paper, we apply the framework of MPS and PEPS to the
classification of gapped quantum phases by studying systems with exact MPS
and PEPS ground states. Here, we define two gapped systems to be in the
same phase if and only if they can be connected by a smooth path of gapped
local Hamiltonians. Along such a path, all physical properties of the
state will change smoothly, and as the system follows a quasi-local
evolution,~\cite{hastings:quasi-adiabatic} global properties are
preserved. A vanishing gap, on the other hand, will usually imply a
discontinuous behavior of the ground state and affect global properties of
the system.  In addition, one can impose symmetries on the Hamiltonian
along the path, which in turn leads to a more refined classification of
phases.

In the presence of symmetries, the above definition of gapped quantum
phases can be naturally generalized to systems with symmetry breaking,
i.e., degenerate ground states, as long as they exhibit a gap above the
ground state subspace. Again, ground states of such systems are well
approximated by MPS and PEPS, which justifies why we study those phases by
considering systems whose ground state subspace is spanned by MPS and
PEPS.  On the other hand, the approach will not work for gapless phases,
as the MPS description typically cannot be applied to them.

\subsection{Results}

Using our framework, we obtain the following classification of quantum phases:

For one-dimensional (1D) gapped systems with a unique ground state, we
find that there is only a single phase, represented by the product state.
Imposing a constrain in form of a local symmetry $U_g$ (with symmetry
group $G\ni g$) on the Hamiltonian leads to a more rich phase diagram. It
can be understood from the way in which the symmetry acts on the virtual
level of the MPS, $U_g\cong V_g\otimes \bar V_g$, where the $V_g$ are
projective representations of $G$. In particular, different phases under a
symmetry are labelled by the different equivalence classes of projective
representations $V_g$ of the symmetry group
$G$,~\cite{pollmann:1d-sym-protection-prb} which are in one-to-one
correspondence to the elements of its second cohomology group
$\mathrm{H}^2(G,\mathrm{U}(1))$ (this has been previously studied in
Ref.~\onlinecite{chen:1d-phases-rg}).

For one-dimensional gapped systems with degenerate ground states, we find
that in the absence of symmetries, all systems with the same ground state
degeneracy  can be transformed into another along a gapped
adiabatic path.  In order to make these degeneracies stable against
perturbations, symmetries need to be imposed on the Hamiltonian. We find
that any such symmetry decomposes into two parts, $P_g$ and $W_h$.  Here,
$P_g$ acts by permuting the symmetry broken ground states of the system.
To describe $W_h$, choose a ``reference'' ground state, and let
$H\subset G$ be the subgroup for which $P_h$ acts trivially on the
reference state: Then, $W_h$ ($h\in H$) is a unitary symmetry of the
reference state, which again acts on the virtual level as $W_h\cong
V_h\otimes \bar V_h$, with $V_h$ a projective representation of $H$.
Together, $P_g$ and $W_h$ form an induced representation.  The different
phases of the system are then labelled both by the permutation action $P_g$ of
$U_g$ on the symmetry broken ground states (or alternatively by the subgroup
$H\subset G$ for which $P_h$ leaves the reference state invariant),
and by the equivalence classes of projective representations $V_h$
of $H$.

Our classification of phases is robust with respect to the definition of
the gap: Two systems which are within the same phase can
be connected by a path of Hamiltonians which is gapped even in the
thermodynamic limit; conversely, along any path interpolating between
systems in different phases the gap closes already for any (large enough)
finite chain.  On the other hand, we demonstrate that the classification
of phases is very sensitive to the way in which the symmetry constraints
are imposed, and we present various alternative definitions, some of which
yield more fine-grained classifications, while others result in the
same classification as without symmetries. In particular, we also find
that phases under symmetries are not stable under taking
multiple copies, and thus should not be regarded a resource in the quantum
information sense.

Parts of our results can be generalized to two dimensions (2D), with the
limitation that we can only prove gaps of parent Hamiltonians associated with
PEPS in restricted regions. These regions include systems with unique
ground states and with local symmetry breaking as well as topological
models. We show that within those regions, these models label different
quantum phases, with the product state, Ising Hamiltonians, and Kitaev's
double models~\cite{kitaev:toriccode} as their representatives. We also
find that in these regions, imposing symmetries on the Hamiltonian does
not alter the phase
diagram, and, more generally, that symmetry constraints on two- and
higher-dimensional systems must arise from a different mechanism than in
one dimension.

As a main tool for our proofs, we introduce a new standard form for MPS and
PEPS which we call the \emph{isometric form}. Isometric forms are
renormalization fixed points which capture the relevant features of the
quantum state under consideration, both for MPS and for the relevant
classes of PEPS. Parent Hamiltonians of MPS can be transformed into their
isometric form along a gapped path, which provides a way to renormalize
the system without actually blocking sites.  This reduces the
classification of quantum phases to the classification of phases for
isometric MPS/PEPS and their parent Hamiltonians, which is considerably
easier to carry out due to its additional structure.

\subsection{Structure of the paper}

The paper is structured as follows: In Section II, we prove the results
for the one-dimensional case: We start by introducing MPS (Sec.~II A) and
parent Hamiltonians~(Sec.~II~B), and define  phases without and with
symmetries (Sec.~II C). We then introduce the isometric form and show that
the problem of classifying phases can be reduced to classifying isometric
forms (Sec.~II~D). Subsequently, we first classify 1D phases without
symmetries~(Sec.~II E), then phases of systems with unique ground states
under symmetries (Sec.~II F), and finally phases of symmetry broken
systems under symmetries~(Sec.~II G).

In Section III, we discuss the 2D scenario. We start by introducing PEPS
(Sec.~III A) and characterize the region in which we can prove a gap
(Sec.~III B).  We then classify PEPS without symmetries (Sec.~III C) and
show that symmetries don't have an effect comparable to one dimension
(Sec.~III D).

Section IV contains discussions of various topics which have been omitted
from the preceding sections.  Most importantly, in Sec.~IV B and C, we
discuss various ways in which phases, in particular in the presence of
symmetries, can be defined, and the way in which this affects the
classification of phases, and in Sec.~\ref{sec:examples} we provide
examples illustrating our classification.

\section{Results in one dimension}

In this section, we will derive the classification of phases for
one-dimensional systems both with unique and with degenerate 
 ground states, and both in the absence and the presence
of symmetries. We will start by giving the necessary definitions---we
will introduce Matrix Product States and their parent Hamiltonians, and
define what we mean by phases both without and with symmetries. Then, we
will state and prove the classification of phases for the various
scenarios (with some technical parts placed in appendices). 

Note that for clarity, we will keep discussions in this section to a
minimum.  Extensive discussion of various aspects (motivation of the
definitions, alternative definitions, etc.) can be found in
Sec.~\ref{sec:discussion}.

\subsection{Matrix Product States}

In the following, we will study translational invariant systems on a
finite chain of length $N$ with periodic boundary conditions. While we do
not consider the thermodynamic limit, we require relevant properties such
as spectral gaps to be uniform in $N$.

\begin{figure}[t]
  \includegraphics[width=4cm]{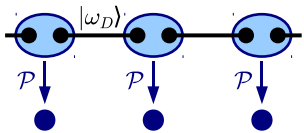}
\caption{\label{fig:mps}
MPS are constructed by applying a linear map $\mc P$ to maximally
entangled pairs $\ket{\omega_D}:=\sum_{i=1}^D \ket{i,i}$ of \emph{bond
dimension} $D$.
}
\end{figure}

\sss{Definition of MPS}%
Consider a spin chain $(\mathbb C^d)^{\otimes N}$.  A
(translational invariant) Matrix Product State (MPS) $\MPS{\mc P}$ of
\emph{bond dimension} $D$ on $(\mathbb C^d)^{\otimes N}$ is constructed by
placing maximally entangled pairs 
\[
\ket{\omega_D}:=\sum_{i=1}^D\ket{i,i}
\]
between adjacent sites and applying a linear map $\mc P:\mathbb C^D\otimes
\mathbb C^D\rightarrow \mathbb C^d$, as depicted in
Fig.~\ref{fig:mps};\cite{verstraete:dmrg-mps} this is, $\MPS{\mc P}=\mc
P^{\otimes N}\ket{\omega_D}^{\otimes N}$. The map $\mc P$ is sometimes
called the \emph{MPS projector} (though it need not be a projector).

\sss{Standard form}%
The definition of MPS is robust under blocking sites: Blocking $k$ sites
into one super-site of dimension $d^k$ gives a new MPS with the same bond
dimension, with projector
$\mc P'=\mc P^{\otimes k}\ket{\omega_D}^{\otimes (k-1)}$, where
the maximally entangled pairs are again placed as in Fig.~\ref{fig:mps}.

By blocking and using gauge degrees of freedom (including changing $D$),
any MPS which is well-defined in the thermodynamic limit can be brought
into its
\emph{standard form}, where $\mc P$ is supported on a ``block-diagonal''
space,
$(\mathrm{ker}\:\mc P)^\perp=\mathcal
H_1\oplus\dots\oplus H_\mc A$, with 
\begin{equation}
        \label{eq:mps-def:Halpha-directsum}
\mc H_\alpha = \mathrm{span}\{\ket{i,j}: \zeta_{\alpha-1}<i,j\le \zeta_\alpha\}\ .
\end{equation}
Here,
$0=\zeta_0<\zeta_1<\dots<\zeta_{\mc A}=D$ gives a partitioning of
$1,\dots,D$; we define $D_i:=\zeta_{i}-\zeta_{i-1}$. The case where $\mc
A=1$ (i.e.\ $\mc P$ is injective) is called the \emph{injective} case, whereas
the case with $\mc A>1$ is called \emph{non-injective}.

In the following, we will assume w.l.o.g.\ that all MPS are in their
standard form---note that this might involve the blocking of sites of the
original model. Moreover, we will impose that $\mc P$ is surjective, this
can be achieved by restricting $\mathbb C^d$ to the image of $\mc P$.

\subsection{Parent Hamiltonians}

For any MPS (in its standard form) one can construct local \emph{parent
Hamiltonians} which have this MPS as their ground state.  A (translational
invariant) parent Hamiltonian 
\[
H=\sum_{i=1}^N h(i,i+1)
\]
consists of local terms
$h(i,i+1)\equiv h\ge0$ acting
on two adjacent sites $(i,i+1)$ whose kernels exactly support the two-site
reduced density operator of the corresponding MPS, i.e.,
\begin{equation}
        \label{eq:mps-def:parentham-def}
\mathop{\mathrm{ker}} h = 
(\mc P\otimes \mc P) (\mathbb C^D\otimes \ket{\omega_D}\otimes
\mathbb C^D)\ .
\end{equation}
By construction, $H\ge0$ and $H\MPS{\mc P}=0$, i.e.,
$\MPS{\mc P}$ is a ground state of $H$. It can be shown that the ground
state space of $H$ is $\mathcal A$-fold degenerate, and spanned by the
states $\MPS{\mc P\vert_{\mc H_\alpha}}$, where
$\mc P\vert_{\mc H_\alpha}$ is the restriction of $\mc P$ to $\mc
H_\alpha$.%
\cite{fannes:FCS,nachtergaele:degen-mps,perez-garcia:mps-reps}
This associates a family of parent Hamiltonians to every MPS, and by
choosing $h$ a projector, the mapping between MPS and
Hamiltonians becomes one-to-one. 

Given this duality between MPS and their parent Hamiltonians, we will use
the notion of MPS and parent Hamiltonians interchangingly whenever
appropriate.

\subsection{Definition of quantum phases}

Vaguely speaking, we will define two systems to be in the same phase if
they can be connected along a continuous path of gapped local
Hamiltonians, possibly preserving certain symmetries; here, \emph{gapped}
means that the Hamiltonian keeps its spectral gap even in the
thermodynamic limit.  The intuition is that along any gapped path, the
expectation value of any local observable will change smoothly, and the
converse is widely assumed to also hold.~\cite{chen:2d-rg}

The rigorous definitions are as follows.

\sss{\label{sec:phases-definition-no-sym}
Phases without symmetries}%

Let $H_0$ and $H_1$ be a family of translational invariant gapped local
Hamiltonians on a ring (with periodic boundary conditions).  Then, we say
that $H_0$ and $H_1$ are in the same phase if and only if there exists a
finite $k$ such that after blocking $k$ sites, $H_0$ and $H_1$ are
two-local,
\[
H_p=\sum_{i=1}^{N} h_p(i,i+1)\ ,\quad p=0,1\ ,
\]
and there exists a translational invariant
path 
\[ 
H_\gamma = \sum_{i=1}^N h_\gamma(i,i+1)\ ,\quad 0\le \gamma\le 1\ ,
\]
with two-local $h_\gamma$ such that
\begin{packed_enum}
\item $h_0=h_{\gamma=0}$\,, $h_1=h_{\gamma=1}$ 
\item $\|h_\gamma\|_\mathrm{op}\le 1$
\item $h_\gamma$ is a continuous function of $\gamma$
\item $H_\gamma$ has a spectral gap above the ground
        state manifold which is bounded below by some constant
        $\Delta>0$ which is independent of $N$ and $\gamma$.
\end{packed_enum}
In other words, two Hamiltonians are in the same phase if they can be
connected by a local, bounded-strength, continuous, and gapped path.

Note that this definition applies both to Hamiltonians with unique and
with degenerate ground states.

\sss{Phases with symmetries
\label{sec:def-sym-phases}}%

Let $H_p$ ($p=0,1$) be a Hamiltonian acting on $\mc H_p^{\otimes N}$, $\mc
H_p=\mathbb C^{d_p}$, and let $U^p_g$ be a linear unitary representation of some
group $G\ni g$ on $\mc H_p$. We then say that $U_g$
is a symmetry of $H_p$ if $[H_p,(U_g^p)^{\otimes N}]=0$ for all $g\in G$;
note that $U_g^p$ is only defined up to a one-dimensional representation
of $G$, $U_g^p\leftrightarrow e^{i\phi_g^p} U_g^p$.
Then, we say that $H_0$ and $H_1$ are in the same phase under the symmetry
$G$ if there exists a phase gauge for $U_g^0$ and $U_g^1$ and a
representation $U=U^0_g\oplus U^1_g\oplus
U^{\mathrm{path}}_g$ of $G$ on a Hilbert space $\mc H = \mc H_0\oplus \mc
H_1 \oplus \mc H_{\mathrm{path}}$, and an interpolating
path $H_\gamma$ on $\mc H$ with the properties given in the preceding
section, such that $[H_\gamma,U_g^{\otimes N}]=0$, and where $H_0$ and
$H_1$ are supported on $\mc H_0$ and $\mc H_1$, respectively.

There are a few points to note about this definition: First, we allow for
an arbitrary representation of the symmetry group along the path; we will
discuss in Sec.~\ref{sec:discussion-sym} why this is not a restriction.
Second, we impose that $H_0$ and $H_1$ are supported on orthogonal Hilbert
spaces: This allows us to compare e.g.\ the spin-$1$ AKLT state with the
spin-$0$ state under $\mathrm{SO(3)}$ symmetry, but we will impose this
even if the two representations are the same; we will discuss how to
circumvent this in Sec.~\ref{sec:discussion-sym}.  Note that, just as
without symmetries, this definition should be understood after an
appropriate blocking of sites.

\sss{Robust definition of phases}

In addition to the properties listed in the definition of phases in
Sec.~\ref{sec:phases-definition-no-sym} above, one usually requires a
phase to be robust, i.e., an open set in the space of allowed
Hamiltonians: For every Hamiltonian
\[
H=\sum_{i=1}^N h(i,i+1)
\]
there should be an $\epsilon>0$ such that 
\[
H=\sum_{i=1}^N \Big[\,h(i,i+1)+\epsilon\, k(i,i+1)\Big]
\]
is in the same phase for any bounded-strength $k(i,i+1)$ which obeys the
required symmetries.

We are not going to rigorously address robustness of phases in the present
paper; however, it should be pointed out that, in the absence of
symmetries, Hamiltonians with degenerate MPS ground states do not satisfy
this property: In its standard form, the different ground states of a
non-injective MPS are \emph{locally} supported on linearly independent
subspaces, and we can use a translational invariant local perturbation
$\epsilon\, k(i)$ to change the energy of any of the ground states
proportionally to $\epsilon N$, thereby closing the gap for a $N\propto
1/\epsilon$. On the other hand, in the presence of a symmetry which
permutes the different ground states (such a symmetry always exists),
those perturbations are forbidden, and the phase becomes stable. (A
rigorous stability proof for MPS phases would make use of the stability
condition for frustration-free Hamiltonians proven in
Ref.~\onlinecite{michalakis:local-tqo-ffree} and its generalization to
symmetry-broken phases analogous to the one discussed in
Ref.~\onlinecite{bravyi:local-tqo-simple}: The condition is trivially
satisfied by the renormalization fixed points of
MPS,\cite{verstraete:renorm-MPS} and using the exponential convergence of
MPS to their fixed point,\cite{verstraete:renorm-MPS} the validity of the
stability condition, and thus the stability of MPS phases, follows.)

Therefore, when classifying phases of systems with degenerate ground
states, one should keep in mind that in order to make this a robust
definition, a symmetry which protects the degeneracy is required.

\sss{Restriction to parent Hamiltonians}%
We want to classify the quantum phases of gapped Hamiltonians which have
exact MPS ground states (or, in the case of degeneracies, the
same ground state subspace as the corresponding parent Hamiltonian).
Fortunately, with our definition of phases it is sufficient to classify
the phases for parent Hamiltonians themselves: Given any two gapped
Hamiltonians $H$ and $H'$ which have the same ground state subspace, the
interpolating path $\gamma H+(1-\gamma)H'$ has all desired properties, and
in particular it is gapped. Note that  this also shows that all parent
Hamiltonians for a given MPS are interchangeable.

\subsection{The isometric form}

\sss{Reduction to a standard form}%
Given two MPS $\MPS{\mc P_p}$, $p=0,1$, together with their nearest-neighbor 
parent Hamiltonians $H_p$, we want to see whether $H_0$ and $H_1$ are in
the same phase, i.e., whether we can construct a gapped interpolating path. We
will do so by interpolating between $\mc P_0$ and $\mc P_1$ along a path
$\mc P_\gamma$, in such a way that it yields a path $H_\gamma$ in the
space of parent Hamiltonians which satisfies the necessary continuity and
gappedness requirements.

\begin{figure}[t]
\includegraphics[width=0.95\columnwidth]{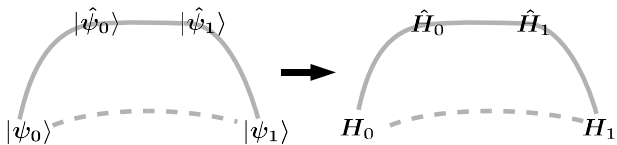}
\caption{\label{fig:inter-path}
Construction of the interpolating path for MPS and parent Hamiltonians.
Instead of interpolating between the MPS $\ket{\psi_0}$ and $\ket{\psi_1}$
directly (dotted line), we first show how to
interpolate each of the two states towards a standard from
$\vert\hat\psi_p\rangle$, the \emph{isometric form}, and then construct an
interpolating path between the isometric forms. Note that using the
parent Hamiltonian formalism, any such path in the space of MPS 
yields a path in the space of Hamiltonians right away.
}
\end{figure}

In order to facilitate this task, we will proceed in two steps, as
illustrated in Fig.~\ref{fig:inter-path}: In a first step, we will
introduce a standard form for each MPS---the \emph{isometric form}---which
is always in the same phase as the MPS itself. This will reduce the task
of classifying phases to the classification of phases for isometric MPS,
which we will pursue subsequently.

\sss{The isometric form}%
Let us now introduce the \emph{isometric form} of an MPS. The isometric
form captures the essential entanglement and long-range properties of the
state and forms a fixed point of a renormalization
procedure,\cite{verstraete:renorm-MPS} and every MPS can be brought into
its isometric form by stochastic local
operations.\cite{bennett:multipart-ent-measures} Most importantly, as we
will show there exists a gapped path in the space of parent Hamiltonians
which interpolates between any MPS and its isometric form.

Given an MPS $\MPS{\mc P}$, decompose 
\begin{equation}
        \label{eq:1d-iso:polardec}
        \mathcal P=Q\: W\ ,
\end{equation}        
with $W$ an isometry $WW^\dagger=\openone$ and $Q>0$, by virtue of a polar
decomposition of $\mc P\vert_{(\mathrm{\ker}\: \mc P)^\perp}$; w.l.o.g.,
we can assume $0<Q\le \openone$ by rescaling $\mc P$. The \emph{isometric
form} of of $\MPS{\mc P}$ is now defined to be $\MPS{W}$, the MPS
described by $W$, the isometric part of the tensor $\mc P$.  

\begin{figure}[t]
\includegraphics[width=.95\columnwidth]{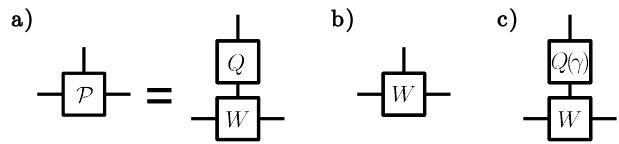}
\caption{\label{fig:iso-form}
Isometric form of an MPS. \textbf{a)} The MPS projector $\mc P$ can be
decomposed into a positive map $Q$ and an isometric map $W$. \textbf{b)}
By removing $Q$, one obtains the isometric form $W$ of the MPS.
\textbf{c)} Interpolation to the isometric form is possible by letting
$Q_\gamma=\gamma Q + (1-\gamma)\openone$.
}
\end{figure}

To see that $\MPS{\mc P}$ and $\MPS{W}$ are in the same phase, define an
interpolating path $\MPS{\mc P_\gamma}$, $\mc P_\gamma = Q_\gamma W$, with
\[
Q_\gamma=\gamma Q+(1-\gamma)\openone\ ,\qquad 1\ge\gamma\ge0\ ;
\]
note that the path can be seen as a stochastic deformation $\MPS{\mc
P_\gamma}=Q_\gamma^{\otimes N}\MPS{\mc P_0}$,
cf.~Fig.~\ref{fig:iso-form}.  Throughout the path, the MPS stays in
standard form, and in the non-injective case, the blocking pattern which
is encoded in the structure of $\mathrm{ker}\: \mc P$ stays unchanged.

Let us now see that $\MPS{\mc P}\equiv\MPS{\mc P_1}$ and its isometric
form $\MPS{W}\equiv\MPS{\mc P_0}$ are in the same phase. To this end,
consider the parent Hamiltonian $H_0=\sum h_0(i,i+1)$ of the isometric MPS
$\MPS{\mc P_0}$, with $h_0$ a projector.
Let 
\[
\Lambda_\gamma=\left(Q_\gamma^{-1}\right)^{\otimes 2}\ ,
\] 
and define the \emph{$\gamma$-deformed Hamiltonian} 
\[
H_\gamma:=\sum h_\gamma(i,i+1)\ ,\mbox{\ with } 
h_\gamma:=\Lambda_\gamma h_0\Lambda_\gamma\ge0\ .
\]
Since $h_0\MPS{\mc P_0}=0$, it follows that $h_\gamma\MPS{\mc P_\gamma}=0$
(and in fact, the kernel of $h$ is always equal to the support of the
two-site reduced state), i.e., $H_\gamma$ is a parent Hamiltonian of
$\MPS{\mc P_\gamma}$.  Note that the family $H_\gamma$ of Hamiltonians is
continuous in~$\gamma$ by construction.

It remains to show that the path $H_\gamma$ is uniformly gapped, i.e.,
there is a $\Delta>0$ which lower bounds the gap of $H_\gamma$ uniformly
in $\gamma$ and the systems size $N$: This establishes that the $\MPS{\mc
P_\gamma}$ are all in the same phase. The derivation is based on a result of
Nachtergaele~\cite{nachtergaele:degen-mps} (extending the results
of Ref.~\onlinecite{fannes:FCS} for the injective case), where a lower bound (uniform
in $N$) on the gap of parent Hamiltonians is derived, and can be found in
Appendix A. The central point is that the bound on the gap depends on the
correlation length $\xi$ and the gap of $H_\gamma$ restricted to $\xi$
sites, and since both depend smoothly on $\gamma$, and $\xi\rightarrow 0$
as $\gamma\rightarrow 0$,  a uniform lower bound on the gap follows; for
the non-injective case, one additionally needs that the overlap of
different ground states goes to zero as $\gamma\rightarrow 0$.

\sss{Isometric form and symmetries}%
An important point about the isometric form $\MPS{\mc P_0} $ of an MPS
$\MPS{\mc P_1}$ is that both remain in the same phase even if symmetries
are imposed. The reason is that using gauge transformations 
\[
\mc P\,\longleftrightarrow\,\mc P\big(\,Y\otimes (Y^{-1})^T\,\big)
\]
---such transformations do not change $\MPS{\mc
P}$\,---any $\mc P$ can be brought into a standard form where 
$\tr_\mathrm{left}[\mc P^\dagger \mc P]=\openone_{\mathrm{right}}$
[cf. Refs.~\onlinecite{fannes:FCS,perez-garcia:mps-reps}];
in this standard form, any symmetry $U_g^{\otimes N}$ of the MPS $\MPS{\mc
P}$ can be understood as some unitary $X_g$ acting on the virtual
level,~\cite{david:stringorder-1d}
\[ 
U_g \mc P = \mc P X_g\ .
\] 
In the polar decomposition $\mc P=QW$, Eq.~(\ref{eq:1d-iso:polardec}),
we have that $Q=\sqrt{\mc P \mc P^\dagger}$, and thus
\begin{linenomath}
\begin{align*}
Q^2=\mc P \mc P^\dagger &= 
     U_g^\dagger\, \mc P \, X_g
     X_g^\dagger \mc P^\dagger \, U_g
\\
    &= U_g^\dagger \mc P \mc P^\dagger U_g
    = U_g^\dagger Q^2 U_g\ .
\end{align*}
\end{linenomath}
This is, for any $g\in G$ the matrices $Q^2$ and $U_g$ are diagonal in a
joint basis, and therefore $[Q,U_g]=0$, and it follows that both the
interpolating path $\MPS{\mc P_\gamma}=Q_\gamma^{\otimes N}\MPS{\mc P_0}$
and its parent Hamiltonian $H_\gamma$ are invariant under $U_g$.

\subsection{Phase diagram without symmetries}

The preceding discussion show thats in order to classify quantum
phases (both without and with symmetries), it is sufficient to consider
isometric MPS and their parent Hamiltonians.  In the following, we will
carry out this classification for the scenario where no symmetries are
imposed. We will find that without symmetries, the phase of the system
only depends on the ground state degeneracy, as any two systems with the
same ground state degeneracy are in the same phase (and clearly the ground
state degeneracy cannot be changed without closing the gap).

\sss{Unique ground state}

\begin{figure}[t]
        \includegraphics[height=1.7cm]{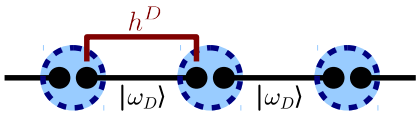}
\caption{\label{fig:iso-injective}
The isometric form of an injective MPS consists of maximally entangled
states $\ket{\omega_D}$ with bond dimension $D$ between adjacent sites,
where each site is a $D^2$--level system.  The local Hamiltonian terms
$h^D$, Eq.~(\ref{eq:phase-nosym:iso-hamiltonian}), acts only on the degrees
of freedom corresponding to the central entangled pair.}
\end{figure}

Let us start with the injective case where 
the Hamiltonian has a unique ground state. In that case, the isometry
$W$ is a unitary. We can now continuously undo the rotation $W$---this
clearly is a smooth gapped path and does not change the phase.  
This yields the state $\MPS{\mc P=\openone}=\ket{\omega_D}^{\otimes N}$,
consisting of maximally entangled pairs of dimension $D$ between
adjacent sites which can only differ in their bond dimension $D$,
cf.~Fig.~\ref{fig:iso-injective}.  The parent Hamiltonian is a sum of
commuting projectors of the form 
\begin{equation}
        \label{eq:phase-nosym:iso-hamiltonian}
h^D=\openone-\ket{\omega_D}\!\bra{\omega_D}
\end{equation}
and thus gapped. (Here, $h^D$ should be understood as acting trivially on
the leftmost and rightmost ancillary particle, and non-trivially on the
middle two, cf.~Fig.~\ref{fig:iso-injective}.)
 Moreover, for different $D$ and $D'$ one can interpolate
between these states via a path of commuting Hamiltonian with local terms
\begin{equation}
        \label{eq:nosym-inj:h-theta}
h^\theta=\openone-\ket{\omega(\theta)}\!\bra{\omega(\theta)}\ ,
\end{equation}
where $\ket{\omega(\theta)}= \sqrt{\theta}\ket{\omega_D}
+\sqrt{1-\theta}\ket{\omega_{D'}}$.  It follows that any two isometric
injective MPS, and thus any two injective MPS, are in the same phase. In
particular, this phase contains the product state as its canonical
representant.

Note that in the previous derivation, the dimensions of the local Hilbert
spaces differ. One can resolve this by thinking of both systems as being
embedded in a larger Hilbert space, or by adding ancillas; we discuss
this further is Sec.~\ref{sec:discussion}.

\sss{Degenerate ground states}

\begin{figure}
\includegraphics[height=1.7cm]{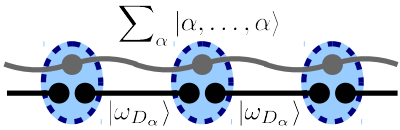}
\caption{\label{fig:ghz-iso-form}
Structure of the isometric form for non-injective MPS: The state consists
of a GHZ state $\sum_\alpha\ket{\alpha,\dots,\alpha}$ (gray) and maximally
entangled pairs between adjacent sites $\ket{\omega_{D_\alpha}}$, where
the bond dimension $D_\alpha$ can couple to the value $\alpha$ of the GHZ
state.  }
\end{figure}

Let us now consider the case of of non-injective MPS, i.e., systems with
degenerate ground states? First, consider the case
with block sizes $D_\alpha:=\dim \mc H_\alpha=1$, i.e., $\mc P\equiv\mc
P_\mathrm{GHZ}=\sum_\alpha \ket{\alpha}\!\bra{\alpha,\alpha}$ (up to
a rotation of the physical system): This
describes an $\mc A$-fold degenerate GHZ state with commuting Hamiltonian
terms 
\begin{equation}
        \label{eq:nosym-noninj:ghz-ham-local}
h=\openone-\sum_\alpha\ket{\alpha,\alpha}\!\bra{\alpha,\alpha}\ .
\end{equation}
For $D_\alpha\ne1$, we
have that (again up to local rotations) 
\[
\mc P=\sum_\alpha\ket{\alpha}\!\bra{\alpha,\alpha}\otimes
\openone_{D_\alpha^2}\ ,
\]
where $\openone_{D_\alpha^2}$ acts on $\mc H_\alpha$ in
(\ref{eq:mps-def:Halpha-directsum}).  This describes a state 
\begin{equation}
        \label{eq:nosym-noninj:most-general-gs}
\sum_\alpha 
    \ket{\alpha,\dots,\alpha}\otimes\ket{\omega_{D_\alpha}}^{\otimes N}
\end{equation}
(cf.~Fig.~\ref{fig:ghz-iso-form}), i.e., a GHZ state with additional local
entanglement between adjacent sites, where the amount of local
entanglement $D_\alpha$ can depend on the value $\alpha$ of the GHZ state.
[Since the $D_\alpha$ can be different, the correct interpretation of 
(\ref{eq:nosym-noninj:most-general-gs})
is $\ket{\omega_{D_1}}^{\otimes N}\oplus\cdots\oplus
\ket{\omega_{D_{\mc A}}}^{\otimes N}$, with the $\ket{\alpha}$ labelling
the direct sum components locally.] The corresponding Hamiltonian is a
sum of the GHZ Hamiltonian (\ref{eq:nosym-noninj:ghz-ham-local})
and terms
\[
\sum_\alpha\ket\alpha\bra\alpha\otimes
(\openone-\vert\omega_{D_\alpha}\rangle\langle\omega_{D_\alpha}\vert)\ ,
\]
which are responsible for the local entanglement. This Hamiltonian
commutes with the GHZ part (this can be understood by the fact that the
GHZ state breaks the local symmetry, i.e., there is a preferred local
basis), and one can again interpolate
to the pure GHZ state with $D'_\alpha=1$ analogously to the injective
case, Eq.~(\ref{eq:nosym-inj:h-theta}).

\sss{Summary: Classification of phases}%
Together, we obtain the following result on the classification of 1D
phases in the absence of symmetries: Any two systems with MPS ground
states with the same ground state degeneracy are in the same phase; the
canonical representant of these phases are the product and the GHZ-type
states, respectively. Since the ground state degeneracy cannot be smoothly
changed without closing the gap, this completes the classification.

\subsection{Phase under symmetries: unique ground state}

Let us now discuss the classification of phases in the presence of
symmetries, as defined in Sec.~\ref{sec:def-sym-phases}. We will in the
following first discuss the case of injective MPS, i.e., systems with
unique ground states. Note that we will consider the symmetry
of the blocked MPS, but as we have argued when defining phases under
symmetries, this does not affect the classification.

An important prerequisite for the subsequent discussion is the observation
that any MPS has a gauge degree of freedom
\begin{equation}
        \label{eq:inj-mps-gauge}
        e^{i\vartheta}\MPS{\mc P}=\MPS{e^{i\phi}\mc P(Y\otimes Y^*)}\ ,
\end{equation}
where $Y$ is right-invertible (i.e., there exists $Y^{-1}$ s.th.\
$YY^{-1}=\openone_D$, but $Y$ need not be square), and $Y^*=(Y^{-1})^T$.
Conversely, it turns out that any two $\mc P$ representing the same state
are related by a gauge transformation (\ref{eq:inj-mps-gauge}).

\sss{Projective representations and the classification of phases}%

Let $U_g$ be a linear unitary representation of a symmetry group $G$.  We
start from the fact that for any $U_g$-invariant MPS $\MPS{\mc P}$ and
parent Hamiltonian, there is a standard form for $\mc P$ and a phase
gauge for $U_g$ such that
\begin{equation}
\label{eq:sym:proj-sym}
U_g\,\mc P =  \mc P \, 
    (V_g\otimes \bar V_g)\ ,
\end{equation}
where the bar denotes the complex conjugate.\cite{david:stringorder-1d}
Here, the $V_g$ form a projective unitary representation of group $G$,
i.e., $V_g V_h = e^{i\omega(g,h)} V_{gh}$;
cf.~Appendix~\ref{app:stform-inj-sym}\ for details.  As $V_g$ appears
together with its complex conjugate in (\ref{eq:sym:proj-sym}), it is only
defined up to a phase, $V_g\,\leftrightarrow\, e^{i\chi_g}V_g$, and thus,
$\omega(g,h)$ is only determined up to the equivalence relation 
\[
\omega(g,h)\sim \omega(g,h) +\chi_{gh}-\chi_g-\chi_h\ \mathrm{mod}\,2\pi\ .
\]
The equivalence classes induced by this relation form a group under
addition (this is, tensor products of representations), which is
isomorphic to the second cohomology group $\mathrm{H}^2(G,\mathrm{U}(1))$ of
$G$ over $\mathrm{U}(1)$; thus, we will also call them \emph{cohomology
classes}.

In the following, we will show that in the presence of symmetries, the
different phases are exactly labelled by the cohomology class of the virtual
realization $V_g$ of the symmetry $U_g$ determined by
Eq.~(\ref{eq:sym:proj-sym}); this result was previously found in
Ref.~\onlinecite{chen:1d-phases-rg}.

\sss{Equality of phases}

Let us first show that MPS with the same cohomology class for $V_g$ in
(\ref{eq:sym:proj-sym}) can be connected by a gapped path.  We will do so
by considering the isometric point, where $\mc P$ is unitary; 
recall that the transformation to the isometric point commutes with the
symmetry, so that (\ref{eq:sym:proj-sym}) still holds.
Then, (\ref{eq:sym:proj-sym}) can be rephrased as 
\[
 \hat U_g =  \mc P^\dagger U_g \mc P = 
     V_g\otimes \bar V_g\ ,
\]
this is, the action of the symmetry can be understood as $\hat U_g =
V_g\otimes \bar V_g$ acting on the virtual system, in a basis
characterized by $\mc P$.

Now consider two MPS with isometric forms $\mc P_0$ and $\mc P_1$, and
symmetries 
\begin{linenomath}
\begin{align*}
\hat U_g^0&= \mc P_0^\dagger U_g^0 \mc P_0 = V^0_g \otimes \bar V^0_g\ ,\\
\hat U_g^1&= \mc P_1^\dagger U_g^1 \mc P_1 = V^1_g \otimes \bar V^1_g\ ,
\end{align*}
\end{linenomath}
where $V^0_g$ and $V^1_g$ are in the same cohomology class. 
We can now interpolate between the two MPS with bond dimensions $D_0$ and
$D_1$ (in the ``convenient basis`` corresponding to $\hat U_g^0$ and $\hat
U_g^1$) along the path $\ket{\omega(\gamma)}^{\otimes N}$, where
\begin{equation}
        \label{eq:sym:omega-gamma}
\ket{\omega(\gamma)} = (1-\gamma) \sum_{i=1}^{D_0}\ket{i,i}
    + \gamma\sum_{i=D_0+1}^{D_1}\ket{i,i}\ ,
\end{equation}
which is an MPS with bond dimension $D_0+D_1$.
Again, the parent
Hamiltonian along this path is commuting and thus gapped, and changes
smoothly with $\gamma$. This path can be understood as a path with
symmetry 
\begin{equation}
        \label{eq:1d-sym:jointsym}
(V^0_g\oplus V^1_g)\otimes (\bar V^0_g\oplus
\bar V^1_g) = \hat U_g^0 \oplus \hat U_g^1 \oplus \hat U^\mathrm{path}_g\
,
\end{equation}
(cf.~Ref.~\onlinecite{chen:1d-phases-rg}), with 
\[
\hat U^\mathrm{path}_g = (V^0_g\otimes \bar V^1_g)
    \oplus (V^1_g\otimes \bar V^0_g)
\]
(this is where equality of the cohomology classes is required, since only
then $\hat U^\mathrm{path}_g$ forms a linear representation).

\subsubsection{Separation of phases
\label{subsubsec:sym-inj-separation}}

As we have seen, MPS for which $V_g$ in Eq.~(\ref{eq:sym:proj-sym}) is in
the same cohomology class fall into the same phase. Let us now show that
conversely, states with different cohomology classes
fall into different phases.  We will prove
this again in the framework of MPS, i.e., we will show that there cannot
be a smooth MPS path connecting two such states.  Note that it is
clear that the interpolation given above cannot work, as now $V^0_g\oplus
V^1_g$ does not form a representation any more.

The idea of the impossibility proof is to consider a chain of arbitrary
length $N$ and show that along any well-behaved
path $H_\gamma$, $\mc P_\gamma$ needs
to change continuously, which results in a continuous change in the way the
symmetry acts on the virtual system. In turn, such a continuous change
cannot change the cohomology class.
While this argument is based on the fact that the chain is
finite (as the continuity bounds depend on $N$), it works for arbitrary system
size $N$; also, our argument implies that in order to interpolate between
two systems with different cohomology classes, i.e., in different phases,
the gap of the Hamiltonian will have to close for a \emph{finite} chain,
and not only in the thermodynamic limit. (This can be understood from the
fact that along an MPS path, the virtual representation of the symmetry is
well defined even for finite chains, so that it cannot change without
closing the gap.  While we believe that cohomology classes label gapped
phases beyond MPS, this will likely not hold exactly for finite chains,
thus leaving the possibility of a higher order phase transition when
interpolating beyond MPS.)
We will now proceed by fixing some $\gamma$ along the path and show
the continuity in an environment $\gamma+\dd\gamma$.

Fix some $\gamma$, with corresponding ground state $\MPS{\mc
P_{\gamma}}$ of $H_{\gamma}$, where $\mc P_{\gamma}$ is w.l.o.g.\ in
its standard form where $U_g\mc P_{\gamma}=\mc
P_{\gamma}(V_g\otimes \bar V_g)$ with $V_g$ unitary.  Now consider
$H_{\gamma+\dd\gamma}$ with $\dd\gamma\ll1$ small, and expand its ground
state as
\[
\MPS{\mc P_{\gamma+\dd\gamma}}=
    \sqrt{1-\lambda^2}\MPS{\mc P_\gamma}
    +\lambda \ket{\chi_\gamma}
\]
with $\langle\chi_\gamma\MPS{\mc P_\gamma}=0$.
With $H_{\gamma+\dd\gamma}=H_\gamma+\dd H$,
\begin{linenomath}
\begin{align}
        \nonumber
0 &= \bra{\bm\mu[\mc P_{\gamma+\dd\gamma}]}
    H_{\gamma+\dd\gamma} \MPS{\mc P_{\gamma+\dd\gamma}} 
\\
        \label{eq:inj-sym:gs-smooth}
& = \lambda^2 \bra{\chi_\gamma}H_\gamma \ket{\chi_\gamma} 
    + \bra{\bm\mu[\mc P_{\gamma+\dd\gamma}]}
    \dd H \MPS{\mc P_{\gamma+\dd\gamma}} 
\\
        \nonumber
&\ge \lambda^2\Delta - \|\dd H\|\ ,
\end{align}
\end{linenomath}
where $\Delta$ is the spectral gap of $H_\gamma$.
Since $\lambda=\big|\MPS{\mc P_\gamma}-\MPS{\mc P_{\gamma+ \dd\gamma}}
\big|$ and $\dd H\rightarrow 0$ as $\dd\gamma\rightarrow 0$, this shows
that $\MPS{\mc P_\gamma}$ is continuous in $\gamma$, and since $\MPS{\mc
P_\gamma}$ is a polynomial in $\mc P_\gamma$, it follows that $\mc
P_\gamma$ can be chosen to be a continuous function of $\gamma$ as well.

Let us now study how the virtual representation of the symmetry is
affected by a continuous change $\mc P_{\gamma+\dd\gamma}= \mc P_\gamma
+\dd\mc P$. Let us first consider the case where $\mc P_\gamma$ and
$\dd\mc P$ are supported on the same virtual space, i.e., the bond
dimension does not change, and let us restrict the discussion to the
relevant space. The representation of the symmetry on the virtual level
becomes\cite{david:stringorder-1d}
\[
Z_g\otimes Z_g^{*} = 
    \mc P_{\gamma+\dd\gamma}^{-1} U_g 
    \mc P_{\gamma+\dd\gamma} \ ,
\]
where  $Z_g\equiv Z_g(\lambda)$ is
invertible, and $Z_g^{*}=(Z_g^{-1})^T$, cf.~Eq.~(\ref{eq:inj-mps-gauge}).
Since both $\mc P_{\gamma+\dd\gamma}$ and its inverse change continuously,
one can find a gauge such that this also holds for $Z_g$.

It remains to see that a continuous change of the representation $Z_g$ does
not change its cohomology class; note that the choice of gauge for $\mc
P_{\gamma+\dd\gamma}$, Eq.~(\ref{eq:inj-mps-gauge}), leads to a transformation $Z_g
\leftrightarrow Y Z_g Y^{-1}$ which does not affect the cohomology class.
Let us first assume that $Z_g\equiv Z_g(\lambda)$ is differentiable, 
and let 
\begin{equation}
        \label{eq:inj:sym:Zgdef}
Z_g = U_g + U_g'\,\dd\gamma\ ,
\end{equation}
and $U_gU_h=e^{i\omega(g,h)}U_{gh}$.
Then, we can start from
\[
Z_g Z_h = e^{i[\omega(g,h)+\omega'(g,h)\dd\gamma]} Z_{gh}
\]
(note that smoothness of $Z_g$ implies smoothness of
$\exp[i\omega(g,h)]=\tr[Z_g Z_h]/\tr[Z_{gh}]$\,)
and substitute (\ref{eq:inj:sym:Zgdef}). Collecting all
first-order terms in $\dd\gamma$, we find that
\[
U_g U_h' + U_g' U_h = 
    e^{i\omega(g,h)} U'_{gh} + i\omega'(g,h) e^{i\omega(g,h)} U_{gh}\ .
\]
Left multiplication with $U_h^{-1} U_g^{-1} = e^{-i\omega(g,h)}
U_{gh}^{-1}$ yields
\[
U_h^{-1}U_h'  + U_h^{-1}U_g^{-1}U_g' U_{h} = 
U_{gh}^{-1} U'_{gh} + i\,\omega'(g,h)\,\openone\ ,
\]
and by taking the trace and using its cyclicity in the second term, we
obtain
\begin{equation}
        \label{eq:1d-sym-inj:change-omega-trivial}
\omega'(g,h) = -i(\phi_g+\phi_h-\phi_{gh})
\end{equation}
with $\phi_g = \tr[U_g^{-1}U_g']$: This proves that differentiable changes of $Z_g$ can
never change the cohomology class of $\omega$. 

In case $Z_g\equiv Z_g(\lambda)$ is continuous but not differentiable, we
can use a smoothing argument: For any $\epsilon$, we can find a
differentiable $\hat Z_g(\lambda,\epsilon)$ such that $\|\hat
Z_g(\lambda,\epsilon)- Z_g(\lambda)\|\le \epsilon$, and define
$\omega_{\lambda,\epsilon}(g,h)$ via
\[
e^{i\omega_{\lambda,\epsilon}(g,h)} = \tr[Z_g(\lambda,\epsilon) 
Z_h(\lambda,\epsilon) Z_{gh}^{-1}(\lambda,\epsilon)]\ .
\]
This $\omega_{\lambda,\epsilon}(g,h)$ (and in particular its real part) 
varies again according to (\ref{eq:1d-sym-inj:change-omega-trivial}) for
any $\epsilon$ and thus does not change its cohomology class, and since 
it is $O(\epsilon)$-close to $\omega_\lambda(g,h)$, the same holds for the
cohomology class of $Z_g(\lambda)$.

In order to complete our proof, we also need to consider the case where
$\mc P_{\gamma+\dd\gamma}$ is supported on a larger space than $\mc
P_\gamma$.
(The converse can be excluded by choosing $\dd\gamma$ sufficiently small.)
This can be done by considering the symmetry on the smaller space, and is
done in Appendix~\ref{app:coh-class-cont-dimchange}. Together, this
continuity argument shows that we cannot change the cohomology class of
the symmetry $U_g$ on the virtual level along a smooth gapped path
$H_\gamma$ and thus completes the classification of phases with unique
ground states in the presence of symmetries.

\subsection{Phases under symmetries: Systems with symmetry breaking}

Having discussed systems with unique ground states, we will now turn our
attention to systems with symmetry breaking, i.e., degenerate ground
states, corresponding to non-injective MPS.  Recall that in that case, the
MPS projector $\mc P$ is supported on a ``block-diagonal'' space 
\begin{equation}
        \label{eq:noninj:diagonalspace}
\mc H=\bigoplus_{\alpha=1}^{\mc A}
\underbrace{\mathbb C^{D_\alpha}\otimes \mathbb C^{D_\alpha}
}_{\displaystyle =:\mc H_\alpha}\ .
\end{equation}

Before starting, let us note that different form the injective case,
symmetry broken systems can be invariant under non-trivial projective
representations as well. However, we can always find a blocking $k$ such
that the symmetry of the symmetry on the blocked system is represented
linearly (see~Sec.~\ref{sec:discussion-sym}), and we will consider that
scenario in the following.

\sss{Induced representations and the structure 
of systems with symmetry breaking}

Let us first explain how the physical symmetry is realized on the
virtual level---we will see that it has the form of a so-called
\emph{induced representation} of a projective representation (the proof
can be found in Appendix~\ref{app:stdform-noninj-sym}). 
Consider an non-injective MPS $\MPS{\mc P}$ and its parent Hamiltonian.
Then, any invariance of the Hamiltonian under a linear unitary
representation of a group $G$ can be understood as invariance under an
equivalent linear representation $U_g$ which---with the correct gauge for
$\mc P$, and in the correct basis---acts on the virtual system as
\begin{equation}
\label{eq:sym:noninj:generalform}
\hat U_g =
P_g
\bigoplus_{\mathfrak a}
\Big(\bigoplus_{\alpha\in \mathfrak a} 
V_h^{\mathfrak a}\otimes \bar V_h^{\mathfrak a}\Big)\ .
\end{equation}
Here, $P_g$ is a permutation representation of $G$ permuting blocks with
different $\alpha$'s in (\ref{eq:noninj:diagonalspace}). $P_g$ leads to a
natural partitioning of $\{1,\dots,\mc A\}$ into minimal subsets
$\mathfrak a$ invariant under the action of all $P_g$ which we call
\emph{irreducible}; the first direct sum in
(\ref{eq:sym:noninj:generalform}) runs over those irreducible sets
$\mathfrak a$.  The $V_h^{\mathfrak a}$ are unitaries, where $h\equiv
h(g,\alpha)$ is a function of $g$ and $\alpha$. Before explaining their
algebraic structure, note that $P_g$ can be thought of as
composed of permutations $P_g^{\mathfrak a}$ acting on irreducible
subsets, and (\ref{eq:sym:noninj:generalform}) can be rewritten as a
direct sum over irreducible subsets,
\begin{equation}
        \label{eq:noninj:sym-irrep-blocks}
\hat U_g=\bigoplus_{\mathfrak a}
P_g^{\mathfrak a}
\Big(\bigoplus_{\alpha\in \mathfrak a} 
    V_h^{\mathfrak a}\otimes \bar V_h^{\mathfrak a}\Big)\ .
\end{equation}

In the following, we will describe the structure of the symmetry for one
irreducible subset $\mathfrak a$; in fact, degeneracies corresponding to
different subsets $\mathfrak a$ are not of particular interest, as they
are not protected by the symmetry (which acts trivially between them) and
are thus not stable under perturbations of the Hamiltonian.  For a single
irreducible subset $\mathfrak a$,  the symmetry 
\begin{equation}
        \label{eq:noninj:sym-oneblock}
P_g^{\mathfrak a}
\Big(\bigoplus_{\alpha\in \mathfrak a} 
V_h^{\mathfrak a}\otimes \bar V_h^{\mathfrak a}\Big)
\end{equation}
has the following structure (which is known as an \emph{induced
representation}): Fix the permutation representation $P_g^{\mf
a}$, pick an element $\alpha_0\in \mf a$, and define a subgroup
$H\subset G$ as 
\[
H=\{g:\pi_g(\alpha_0)=\alpha_0\}
\]
where $\pi_g$ is the action of $P_g$ on the sectors, 
$\mc H_{\pi_g(\alpha)}=P_g \mc H_\alpha$. Further, fix a projective
representation $V_h^{\mf a}$ of the subgroup $H$. Then, the action of
$V_h^{\mf a}$ can be boosted to the full group $G$ in
(\ref{eq:noninj:sym-oneblock}) by picking representatives $k_\beta$ of the
disjoint cosets $k_\beta H$, and labelling them such that
$\pi_{k_\beta}(\alpha_0)=\beta$. Then, for every $g$ and $\alpha$, there
exist unique $h$ and $\beta$ such that
\begin{equation}
        \label{eq:noninj:gbeta-to-kgamma}
gk_\alpha=k_\beta h\ ,
\end{equation}
and this is how $h\equiv h(g,\alpha)$ in (\ref{eq:noninj:sym-oneblock}) is
determined. Note that the action of the permutation is to map $\alpha$ to
$\beta$, so that (\ref{eq:noninj:gbeta-to-kgamma}) carries the full
information of how to boost the representation $V_h^{\mf a}$ of the subset
$H$ to the full group; this is known as an \emph{induced representation}.
(It is straightforward to check that this is well-defined.)

Before classifying the phases, let us briefly comment on the structure of
(\ref{eq:noninj:sym-irrep-blocks}). The sectors $\alpha$ correspond to
different symmetry broken ground states. The splitting into different
irreducible blocks $\mf a$ corresponds to the breaking of symmetries
\emph{not} contained in $U_g$, and it is thus not stable under
perturbations. Within each block, $U_g$ has subsymmetries which can be
broken by the ground states---the $\alpha\in\mf a$---and subsymmetries
which are not broken by the ground states---corresponding to the symmetry
action $V_h^{\mf a}\otimes \bar V_h^{\mf a}$ defined on subspaces $H$,
where the symmetry acts as in the injective case.

\sss{Structure of symmetry broken phases}%
In the following, we will prove that two Hamiltonians with symmetry
breaking are in the same phase iff \emph{i)} the permutation
representations $P_g$ are the same (up to relabelling of blocks), and
\emph{ii)} for each irreducible subset $\mf a$, the projective
representation $V_h^{\mf a}$ has the same cohomology class.
(In different words, we claim that two
systems are in the same phase if $U_g$ permutes the symmetry broken ground
states in the same way, and if the effective action of the symmetry on each
symmetry broken ground state satisfies the same condition as in the
injective case.) Note that having the same $P_g$ allows us to even
meaningfully compare the $V_h^{\mf a}$ as the subgroups $H$ can be chosen
equal. Also note that since the permutation is effectively encoded in the
subgroup $H$, we can rephrase the above classification by saying that a
phase is characterized by the choice of a subgroup $H$ together with one of its
cohomology classes.

As a simple example, the Ising
Hamiltonians $H_x=\sum \sigma^x_i\sigma^x_{i+1}$ and 
$H_z=\sum \sigma^z_i\sigma^z_{i+1}$ are in different phases if the $\mathbb
Z_2$ symmetry $(\sigma^x)^{\otimes N}$ is imposed: While $H_z$ can break 
the symmetry, $H_x$ can not and thus belongs to a different phase.

\sss{Equality of phases}%
The proof of equality requires again to devise an interpolating path in
the space of MPS which is sufficiently well-behaved (i.e., it yields a
smooth and gapped path in the set of parent Hamiltonians). It turns out
that we can use essentially the same construction as in the injective case: 
We interpolate along a path
\[
\ket{\omega_1(\gamma)}^{\otimes N}\oplus\dots
\oplus\ket{\omega_{\mc A}(\gamma)}^{\otimes N}
\equiv \sum_\alpha \ket{\alpha,\dots,\alpha}
\ket{\omega_\alpha(\gamma)}^{\otimes N}
\]
in a space with local dimensions $\sum_\alpha(D_\alpha^0+D_\alpha^1)^2$, 
where
\[
\ket{\omega_\alpha(\gamma)} = 
(1-\gamma)\sum_{i=1}^{D_\alpha^0}\ket{i,i}+
(1-\gamma)\sum_{i=D_\alpha^0+1}^{D_\alpha^0+D_\alpha^1}\ket{i,i}\ ;
\]
the joint symmetry is given by
\[
\bigoplus_{\mathfrak a}
P_g^{\mathfrak a}
\Big(\bigoplus_{\alpha\in \mathfrak a} 
(V_h^{\mathfrak a}\oplus W_h^{\mathfrak a})\otimes 
(\bar V_h^{\mathfrak a}\oplus \bar W_h^{\mathfrak a})\Big)\ ,
\]
where $V_h^{\mf a}$ and $W_h^{\mf a}$ denote the projective
representations for the two systems (as in the injective case, it can be
thought of being embedded in the physical symmetry by blocking two sites).
Again, the Hamiltonian along the path is the sum of the GHZ Hamiltonian
and projectors of the form
$\openone-\ket{\omega_\alpha(\gamma)}\bra{\omega_\alpha(\gamma)}$ which
couple to the GHZ degree of freedom; the resulting Hamiltonian is
commuting and therefore gapped throughout the path.

\sss{Separation of phases}%
Let us now show that two phases which differ in either the permutation
representation $P_g$ or the cohomology classes of the $V_h^{\mf a}$
cannot be transformed into each other along a gapped path. As in the
injective case, we will make use of a continuity argument. The argument
will go in two parts: On the one hand, continuity implies that each of the
symmetry broken ground states changes continuously, and thus the
permutation action $P_g$ of $U_g$ stays the same. The effective action
$ V_h^{\mf a}\otimes \bar V_h^{\mf a}$ (modulo
permutation) of the symmetry
on each of the symmetry broken ground states, on the other hand, can be
classified by reducing the problem to the injective case.

Let us first show that the symmetry broken ground states change
continuously.  Let $\ket{\psi^\alpha_\gamma}:=\MPS{\mc P^\alpha_\gamma}$ be
the (orthogonal) symmetry broken ground states of $H_\gamma$, with $\mc
P_\gamma^\alpha:=\mc P_\gamma\vert_{\mc H_\alpha}$ the restriction of $\mc
P_\gamma$ to $\mc H_\alpha$ (this describes an injective MPS). For small
changes $\dd\gamma$, we can again (as in
Sec.~\ref{subsubsec:sym-inj-separation}) use continuity and gappedness of
$H_\gamma$ to show that the ground state subspaces of $H_\gamma$ and
$H_{\gamma+\dd\gamma}=H_\gamma+\dd H$ are close to each other, i.e., there
exist ground states $\ket{\chi^\alpha_{\gamma+\dd\gamma}}$ of
$H_{\gamma+\dd\gamma}$ such that
\[
\big|
    \ket{\psi^{\alpha}_\gamma}-
    \ket{\chi^\alpha_{\gamma+\dd\gamma}}
\big| \le O^*(\dd H)\ ,
\]
where $O^*(\dd H)$ goes to zero as $\dd H$ goes to zero.
Since the $\ket{\chi^\alpha_{\gamma+\dd\gamma}}$ can be expanded in terms of
the $\ket{\psi^\alpha_{\gamma+\dd\gamma}}$, they are MPS,
$\ket{\chi^\alpha_{\gamma+\dd\gamma}}=
\MPS{\mc Q_{\gamma+\dd\gamma}^\alpha}$.   Using continuity of the
roots of polynomials, we infer that we can choose 
$|\mc Q_{\gamma+\dd\gamma}^\alpha-\mc P^\alpha_\gamma|\le O^*(\dd H)$.
On the other hand, this implies that the $\mc
Q_{\gamma+\dd\gamma}^\alpha$
 are almost supported on $\mc H_\alpha$, and thus 
$|\mc Q_{\gamma+\dd\gamma}^\alpha-\mc P^\alpha_{\gamma+\dd\gamma}|\le
O^*(\dd H)$. Together, this shows that
\[
\big|
    \mc P_{\gamma}^\alpha-
    \mc P^\alpha_{\gamma+\dd\gamma}
\big|\le O^*(\dd H)\ ,
\]
i.e., the $\mc P^\alpha_\gamma$, and thus the symmetry broken ground
states $\ket{\psi^\alpha_\gamma}$, change continuously.  Since $P_g$
describes the permutation action of the physical symmetry $U_g$ on the
symmetry broken ground states, which is a discrete representation, it follows
that $P_g$ is independent of $\gamma$, i.e., it cannot be changed along a
gapped path $H_\gamma$.

In a second step, we can now break the problem down to the injective
scenario. To this end, do the following for each irreducible block of 
$P_g$: Fix the $\alpha_0$ used to define the subgroup $H$ and its
representation $V_h^{\mf a}$, restrict the physical symmetry $U_g$ to
$g\in H$, and consider the injective ground state $\MPS{\mc
P^{\alpha_0}_\gamma}$ and the correspondingly restricted parent Hamiltonian
(this can be done by adding local projectors restricting the system to the
subspace given by $\alpha_0$). Since $U_g$, $g\in H$, leaves $\MPS{\mc
P^\alpha_\gamma}$ invariant up to a phase, it is a symmetry of the
restricted parent Hamiltonian; it acts on the virtual level as
$V^{\mf a}_h\otimes \bar V^{\mf a}_h$.  The results for
the injective case now imply that it is impossible to change the
cohomology class of $V^{\mf a}_h$, thus completing the proof.

\section{Two dimensions}

\subsection{Projected Entangled Pair States
\label{sec:2d-peps}}

\subsubsection{Definition}

\begin{figure}[b]
\includegraphics[width=3.5cm]{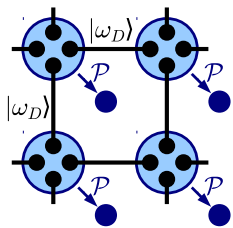}
\caption{\label{fig:peps}
PEPS are constructed analogously to MPS by applying linear maps $\mc P$ to 
a 2D grid of maximally entangled states $\ket{\omega_D}$.
}
\end{figure}

Projected Entangled Pair States (PEPS) form the natural generalization of
Matrix Product States to two dimensions:\cite{verstraete:2d-dmrg}
For $\mc P: (\mathbb
C^D)^{\otimes 4} \rightarrow \mathbb C^d$, the PEPS $\MPS{\mc P}$ is
obtained by placing maximally entangled pairs $\ket{\omega_D}$ on the
links of a 2D lattice and applying $\mc P$ as in
Fig.~\ref{fig:peps}.  As with MPS, PEPS can be redefined by
blocking, which allows to obtain standard forms for $\mc P$, discussed
later on.  Parent Hamiltonians for PEPS are constructed (as in 1D) as sums
of local terms which have the space supporting the $2\times 2$ site
reduced state as their kernel.

\subsubsection{Cases of interest}

As in 1D, each PEPS has an isometric form to which it can be continuously
deformed, yielding a continuous path of $\gamma$-deformed Hamiltonians
along which the ground state degeneracy is preserved. 
There are three
classes of PEPS which are of special interest. 

First, the
\emph{injective case}, where $\mathcal P$ is injective,
and $\MPS{\mc P}$ is the unique ground state of its parent
Hamiltonian.\cite{perez-garcia:parent-ham-2d}

Second, the block-diagonal case, where $(\mathrm{ker}\:\mc
P)^\perp=\bigoplus_{\alpha=1}^{\mc A} \mc H_\alpha$, with $\mc H_\alpha =
\mathrm{span}\{\ket{i,j,k,l}: \zeta_{\alpha-1}<i,j,k,l\le \zeta_\alpha\}$;
this corresponds again to GHZ-type states and Hamiltonians with $\mc
A$-fold degenerate ground states.  These systems are closely related to
the 1D non-injective case---they exhibit breaking of some local symmetry,
and the ground state subspace is spanned by $\MPS{\mc P\vert_{\mc
H_\alpha}}$.

Third, the case where the isometric
form of $\mc P$ is 
\begin{equation}
\label{eq:topo-proj}
\mc P=\sum_g V_g\otimes \bar V_g\otimes W_g \otimes \bar W_g
\end{equation}
(the ordering of the systems is top--down--left--right),
with $V_g$ and $W_g$ unitary representations of a finite group $G$ containing all
irreps of $G$ at least once; this scenario corresponds to systems where
the ground state degeneracy depends on the topology of the system, and
which thus exhibit some form of topological order;\cite{schuch:peps-sym}
in particular, for $V_g$ and $W_g$ the regular representation of $G$, 
the isometric form of these PEPS describes Kitaev's double model of the
underlying group.\cite{kitaev:toriccode} All these three classes have
parent Hamiltonians at the isometric point which are commuting and thus
gapped.
 
\subsection{\label{sec:2d-gap}
Gap in two dimensions}

\subsubsection{Gap in the thermodynamic limit}

The major difference to the case of 1D systems is that it is much more
difficult to assess whether the parent Hamiltonian is gapped in the
thermodynamic limit, and examples which become gapless at some finite
deformation $0<\gamma_\mathrm{crit}<1$ of the isometric form exist. For
instance, the coherent state corresponding to the classical Ising
model\cite{verstraete:comp-power-of-peps} on a hexagonal lattice at the
critical temperature has critical correlations and is thus
gapless,\cite{hastings:gap-and-expdecay} while it is injective and
therefore
its isometric form is gapped.  In fact, this example illustrates that a
smooth change in $\mc P$ can even lead to a non-local change in the PEPS
$\MPS{P}$.

Fortunately, it turns out that in some environment of commuting
Hamiltonians (and in particular in some environment of the three classes
introduced above), a spectral gap can be proven.  To this end,
let $\tilde H=\sum \tilde h_i$, $\tilde h_i\ge\openone$ with ground state
energy $\lambda_{\min}(\tilde H)=0$, where the condition
\begin{equation}
        \label{eq:comm-lowerbnd-for-gap}
\tilde h_i \tilde h_j+\tilde h_j \tilde h_i
\ge -\tfrac18(1-\Delta) (\tilde h_i+\tilde h_j)
\end{equation}
holds for some $\Delta>0$ (here, each $h_i$ acts on $2\times2$ plaquettes
on a square lattice); in particular, this is the case for commuting
Hamiltonians. Then, 
\begin{equation}
        \label{eq:comm-gap-simple}
\tilde H^2=
\sum_i \underbrace{\tilde h_i^2}_{\ge h_i} + 
\sum_{<ij>} \tilde h_i \tilde h_j +
{\sum}^\prime \underbrace{\tilde h_i \tilde h_j}_{\ge0}
\ge \Delta \tilde H
\ ,
\end{equation}
(where the second and third sum run over overlapping and non-overlapping
$\tilde h_i$, $\tilde h_j$, respectively), which implies that $\tilde H$
has a spectral gap between $0$ and $\Delta$, cf.~Ref.~\onlinecite{fannes:FCS}.

As we show in detail in Appendix \ref{appendix:gap}, condition
\eqref{eq:comm-lowerbnd-for-gap} is robust with respect to
$\gamma$-deformations of the Hamiltonian.  In
particular, for any PEPS $\MPS{\mc P}$ with commuting parent Hamiltonian
(such as the three cases presented above), it still holds for the parent
of $Q^{\otimes N}\MPS{\mc P}$ as long as
$\lambda_{\min{}}(Q)/\lambda_{\max{}}(Q)\gtrapprox 0.967$. Thus, while
considering the isometric cases does not allow us to classify all
Hamiltonians as in 1D, we can still do so for a non-trivial subset in the
space of Hamiltonians.

\subsection{Classification of isometric PEPS without symmetries}

Let us now classify the three types of isometric PEPS introduced
previously in the absence of symmetries; together with the results of the
previous subsection, this will provide us with a classification of quantum
phases in some environment of these cases.

\subsubsection{Systems with unique or GHZ-type ground state}

As in one dimension, any injective isometric $\mc P$ can be locally
rotated to the scenario where $\mc P=\openone$, this is, it consists of maximally
entangled pairs between adjacent sites. This entanglement can again be
removed along a commuting path $h^\theta$,
Eq.~(\ref{eq:nosym-inj:h-theta}), as in one dimension.  This implies that
any injective PEPS and its parent Hamiltonian which is sufficiently close
to being isometric is in the same phase as the product state.

The case with block-diagonal $\mc P$, such as for GHZ-like states, is
also in complete analogy to one dimension: Up to a rotation, it is
equivalent to the $\mc A$-fold degenerate GHZ state with additional
maximally entangled pairs between adjacent sites, whose bond dimension
$D_\alpha$ can again couple to the (classical) value of the GHZ state.
This local entanglement can again be removed along a commuting path, as in
one dimension, and we find that all block-diagonal PEPS which are close
enough to being isometric can be transformed into the $\mc A$--fold
degenerate GHZ state along a gapped adiabatic path.

\subsubsection{Systems with topological order}

What about the topological case of Eq.~\eqref{eq:topo-proj}?  Of course,
additional local entanglement $\ket{\omega_D}$ can be present
independently of the topological part of the state, which corresponds to
replacing $V_g$ by $V_g\otimes \openone$ (and correspondingly for $W_g$),
and this entanglement can be manipulated and removed along a commuting
path.  

However, it turns out that the bond dimension $D$ of the local
entanglement can couple to the topological part of the state 
even though there is no local symmetry breaking. 
In particular, the bond dimension $D_\alpha$ can couple to the irreps
$R^\alpha(g)$ of $V_g$ and $W_g$, i.e., we can change the multiplicity
$D_\alpha$ of individual irreps $R^\alpha(g)$, 
\[
\bigoplus_\alpha R^\alpha(g)\;\longleftrightarrow\;
\bigoplus_\alpha R^\alpha(g)\otimes\openone_{D_\alpha}\ .
\]
The interpolation between different multiplicities ${D_\alpha}$ can be
done within the set of commuting Hamiltonians, by observing that the
Hamiltonian consists of two commuting parts:\cite{schuch:peps-sym} One
ensures that the product of each
irrep around a plaquette is the identity, and the other controls the
relative weight of the different subspaces and thus allows to change
multiplicities. The underlying idea can be understood most easily by
considering a two-qubit toy model consisting of the two commuting terms
\begin{linenomath}
\begin{align*}
h_z&=\tfrac12(\openone-Z\otimes Z)\ ,\\
h_x(\theta)&=
    \Lambda_\theta^{\otimes 2} \:
    \tfrac12(\openone-X\otimes X)\:
    \Lambda_\theta^{\otimes 2}\ ,
    \quad
\end{align*}
\end{linenomath}
where $\Lambda_\theta=
\left(\begin{smallmatrix}\theta&0\\0&1\end{smallmatrix}\right)$:
The term $h_z$ enforces the even-parity subspace
$\alpha\ket{00}+\beta\ket{11}$, while $h_x(\theta)$ takes care that the
relative weight within this subspace is $\ket{00}+\theta^2\ket{11}$, which
allows to smoothly interpolate between $\ket{00}$ and $\ket{00}+\ket{11}$
within the set of commuting Hamiltonians.  

Together,  this proves that for a given group $G$, all PEPS of the form
(\ref{eq:topo-proj}), with
representations $V_g$ and $W_g$ which contain all irreducible
representations of $G$, yield PEPS which are in the same phase.  
On the other hand, it is not clear whether the converse holds:
Given two finite groups $G$, $H$ with corresponding representations $V_g$,
$W_g$ and $V_h'$, $W_h'$, for which Eq.~\eqref{eq:topo-proj} yields the
same (or a locally equivalent) map $\mc P$---which means that the two
models are in the same phase---is it true that
the two groups are equal? While we cannot answer this question, let us
remark that since both models can be connected by a gapped path, one can
use quasi-adiabatic
continuation~\cite{hastings:quasi-adiabatic,bachmann:quasi-adiabatic} to
show that their excitations need to have the same braiding statistics;
this is,
the representations of their doubles need to be isomorphic as braided
tensor categories. Note that in
Ref.~\onlinecite{buerschaper:doubles-to-stringnets}, the map $\mc P$ is
used to map doubles to equivalent string-net models.

\subsubsection{More types of local entanglement}

Let us remark that while we have characterized the equivalence classes of
isometric PEPS for the three aforementioned classes, this characterization
is not complete, even beyond the difficulty of proving a gap: There are
PEPS which can be transformed to those cases by local unitaries or
low-depth local circuits, yet $\mc
P$ has a different structure. The reason is that unlike in
1D, local entanglement need not be bipartite. E.g., one could add 
four-partite GHZ states around plaquettes: while this is certainly
locally equivalent to the original state, it will change the kernel of
$\mc P$, since only bipartite maximally entangled states can be described
by a mapping $\mc P\rightarrow \mc P\otimes \openone$. Thus, the previous
classification can be extended to a much larger class of isometric
tensors, by including all symmetries of $\mathrm{ker}\:\mc P$ which can
arise due to adding local entanglement.

\subsection{Symmetries in two dimensions
\label{sec:2d-sym}}

How does the situation change when we impose symmetries on the system and
require the Hamiltonian path to commute with some unitary representation
$U_g$? Surprisingly, imposing symmetries in two dimensions has a much
weaker effect than in one dimension, as we will show in the following. In
particular, we will demonstrate how to interpolate along a
symmetry-preserving path between arbitrary injective PEPS,  and between
any two GHZ-type PEPS given that the permutation action of the symmetry on
the symmetry broken ground states is the same; note that the symmetry can
in particular stabilize the degeneracy of GHZ--type states. Recall,
however, that in the following we only show how to construct continuous
paths of PEPS; in order to turn this into a classification of phases under
symmetries, we need to restrict to the regions characterized in
Sec.~\ref{sec:2d-gap} where we can prove a gap.  Yet, the following
arguments show that the reasoning used for one-dimensional systems with
symmetries will not apply in two dimensions, and a more refined framework
might be needed.

\subsubsection{Systems with unique ground states}

Let us start by studying the injective case. There, it has be
shown~\cite{perez-garcia:inj-peps-syms} that
any two maps $\mc P$ and $\mc P'$ which describe the same PEPS are related
via a gauge transformation
\begin{equation}
        \label{eq:2D-sym:inj-gauge-transform}
\mc P' = \mc P(e^{i\phi}\,Y\otimes Y^*\otimes Z \otimes Z^*)\ ,
\end{equation}
with $Y^*=(Y^{-1})^T$. This implies that any unitary invariance of 
$\MPS{\mc P}$ can be understood as a symmetry 
\[
\hat U_g = \mc P^{-1} U_g \mc P
    =V_g\otimes \bar V_g\otimes W_g\otimes \bar W_g
\]
acting on the on the virtual system, with $V_g$ and $W_g$ projective
unitary representations. 

While this is in complete analogy to the one-dimensional case, there is an
essential difference: The representation of the symmetry on the virtual
level is not invariant under blocking sites. By blocking $k\times \ell$
sites, we obtain a new PEPS projector with symmetries $V_g'=V^{\otimes
k}_g$ and $W_g'=W_g^{\otimes \ell}$, respectively. However, taking tensor
products changes the cohomology class, and in particular, for
any finite-dimensional representation $V_g$ there exists a finite $k$ such
that $V_g^{\otimes k}$ is in the trivial cohomology class. This is, by
blocking a \emph{finite} number of sites, any two PEPS $\MPS{\mc P_0}$ and
$\MPS{\mc P_1}$ can be brought into a form where the symmetry on the
virtual level is represented by $V_g^0$, $W_g^0$, and $V_g^1$, $W_g^1$
which are all in the trivial cohomology class. 

At this point, we can proceed as in the 1D case and construct an
interpolating path which preserves the symmetry using
non-maximally entangled states $\ket{\omega(\gamma)}$
[Eq.~(\ref{eq:sym:omega-gamma})], now with joint symmetry
\[
(V_g^0\oplus V_g^1)\otimes
(\bar V_g^0\oplus\bar V_g^1)\otimes
(W_g^0\oplus W_g^1)\otimes
(\bar W_g^0\oplus\bar W_g^1)\ .
\]
Note that the whole construction can be understood as the sequential
application of two one-dimensional interpolations (since at the isometric
point the horizontal and vertical directions decouple) and thus, all
arguments concerning technical points such as the embedding in the
physical space can be directly transferred.

\subsubsection{Systems with local symmetry breaking}

The results of Ref.~\onlinecite{perez-garcia:inj-peps-syms} for the
relation of two maps $\mc P$ and $\mc P'$,
Eq.~(\ref{eq:2D-sym:inj-gauge-transform}), can be readily generalized to
relate different representations of a PEPS with local symmetry breaking,
i.e., GHZ-type states.  It follows that for any such PEPS, a symmetry can
be understood at the virtual level as
\[
\hat U_g =
P_g
\bigoplus_{\mathfrak a}
\Big(\bigoplus_{\alpha\in \mathfrak a} 
V_h^{\mathfrak a}\otimes \bar V_h^{\mathfrak a}\otimes
W_h^{\mathfrak a}\otimes \bar W_h^{\mathfrak a}\Big)\ .
\]
where again the $P_g$ permutes the sectors, the $\mf a$ are minimal
subsets invariant under $P_g$, and the $V_h^{\mf a}$ and $W_h^{\mf a}$,
together with $P_g$, form an induced projective representation. 

The permutation action $P_g$ of the symmetry is invariant under blocking.
Thus, we can apply the same type of continuity argument as in 1D to show
that different permutations $P_g$ label different phases. (Since this
argument proves that there cannot be a gapped path of Hamiltonians on any
finite chain connecting systems with different $P_g$, it holds
independently of whether we can prove a gap on the Hamiltonian along the
path, as long as the initial and final systems are gapped.)  On the
other hand, the projective representations behave under blocking just as
in the injective case: They map to tensor products of themselves, and
thus, we can choose a blocking such that the $V_h^{\mf a}$ and $W_h^{\mf
a}$ are all in the trivial cohomology class. If moreover the $P_g$ are
equal, we can construct an interpolating path of PEPS just as in one
dimension.

\section{Discussion
\label{sec:discussion}}

In this section, we discuss various aspects which have been omitted in the
previous sections.

\subsection{Matrix Product States}

\subsubsection{Injectivity and translational invariance}

In order to obtain injectivity, it is necessary to block sites (say, $k$ sites
per block). Thus, the notion of locality changes: For instance, a
two-local path of parent Hamiltonians is $2k$-local on the unblocked
system. Also, along such a path we can maintain translational
invariance only under translations by $k$ sites, i.e., on the blocked
system.

However, the number of sites which need to be blocked to obtain
injectivity is fairly small, namely $O(\log D)$ for typical
cases~\cite{perez-garcia:fractionalization} and
$O(D^4)$ in the worst case.~\cite{sanz:wielandt}  In
particular, this is much more favorable than what is obtained using
renormalization methods,\cite{chen:1d-phases-rg} where in addition to
injectivity one has to go to block sizes beyond the correlation
length.

It should be noted that several of our arguments apply without blocking,
thus strictly preserving translational invariance.  In particular, the way
in which symmetries act on the virtual level is the same without blocking,
and the impossibility proofs for interpolating paths also apply equally.
On the other hand, it is not clear whether we can construct interpolating
paths without reaching injectivity of $\mc P$: While we can move to
isometric $\mc P$ along gapped paths, the structure of those isometric
points is now much more rich; for instance, the AKLT projector is of that
form. Also, certain alternative definitions of phases under symmetries
(cf.~Sec.~\ref{sec:discussion:def-of-phases}) behave differently under
translational invariance.  Finally, in the case of symmetry broken
systems,
imposing translational invariance can result in
\emph{projective} representations as physical symmetries (this is
discussed in Sec.~\ref{sec:discussion-sym}), which leads to a much more
involved structure of the 1D induced representation in
Eq.~(\ref{eq:app-noninj-sym:diagonal-part-irrep}).

\subsubsection{Different MPS definition
\label{sec:disc:mps-with-matrices}}

Typically, MPS are defined using a set of matrices $A^i$ as
\[
\sum_{i_1,\dots,i_N}\tr[A^{i_1}\cdots A^{i_N}]\ket{i_1,\dots,i_N}\ .
\]
This definition can be easily related to our definition in terms of MPS
projectors via
\[
\mc P = \sum_{i,\alpha,\beta} A^i_{\alpha\beta}
    \ket{i}\bra{\alpha,\beta}\ .
\]

Injectivity of $\mc P$ translates to the fact that the $\{A^i\}_i$ span the
whole space of matrices, and the ``block-diagonal'' support space in the
non-injective case corresponds to restricting  the $A^i$'s to be
block-diagonal and spanning the space of block-diagonal matrices.

The effect of symmetries in this language can be written as follows:
Eq.~(\ref{eq:sym:proj-sym}) becomes
\[
\sum_j(U_g)_{ij} A^j = V_g^T A^i \bar V_g\ ,
\]
and Eq.~(\ref{eq:sym:noninj:generalform}) 
\[
\sum_j(\hat U_g)_{ij} A^j = P_g^\dagger V_g^T A^i \bar V_gP_g\ ,
\]
where the $P_g$ now permute the blocks of the $A^i$'s, and together with
$V_g$ form an induced representation.

While the matrix formalism using the $A^i$'s is more common,
we choose the projector formulation since we
believe it is more suitable for the purposes of this paper (with the
exception of describing the block structure in the non-injective case):
the $\mc P$, or parts of it, are the maps which we use to conjugate the
Hamiltonian with to get to the isometric form, and which we conjugate the
$U_g$ with to obtain the effective action of the symmetry on the bonds.
Also, in this formulation the isometric point is characterized simply by
maximally entangled pairs with $U\otimes \bar U$ symmetry, rather than by
an MPS with Kronecker delta tensors, and is thus more intuitive to deal
with.

\subsection{Definition of phases
\label{sec:discussion:def-of-phases}}

\subsubsection{Different definitions of phases}

There are other definitions of quantum phases: For instance, instead of
gappedness one can ask for a path of Hamiltonians along which the ground
states do not change abruptly. One can also right away consider the ground
states instead of the Hamiltonian and ask whether two states can be
transformed into each other using (approximately) local
transformations.~\cite{chen:1d-phases-rg}
Both these definitions are implied by ours: The existence of a gapped path
implies that the ground states change smoothly,
cf.~Eq.~(\ref{eq:inj-sym:gs-smooth}), and using quasi-adiabatic
continuation,\cite{hastings:quasi-adiabatic} any gapped path of
Hamiltonians yields a quasi-local transformation between the ground
states.\cite{bachmann:quasi-adiabatic}

\subsubsection{Local dimension and ancillas
\label{app:phasedef-ancillas}}

In our definition of phases, we allowed to compare systems with different
local Hilbert space dimension. One way to think of this is to consider the
smaller system as being embedded in the larger system. A more flexible way is to
allow for the use of ancillas to extend the local Hilbert space. In fact,
these ancillas are automatically obtained when blocking: Recall that we
restricted our attention to the subspace actually used by $\mc P$; the
remaining degrees of freedom (if sufficiently many to allow for a 
tensor product structure) can be used to construct ancillas. An explicit
way to do so is to block two isometric tensors together: The state now
contains a maximally entangled pair $\ket{\omega_D}$ (correlated to the
GHZ in the non-injective case) which can be considered as a
$D^2$--dimensional ancilla system.

Note that we can, in the same way, obtain ancillas for systems with
symmetries: After blocking three isometric sites, the maximally entangled
states in the middle are invariant under $\bar V_g\otimes V_g\otimes\bar
V_g\otimes V_g$. Since also two maximally entanged states between sites
$(1,4)$ and $(2,3)$ are invariant under that symmetry, the symmetry acts
trivially on this two-dimensional subspace which thus constitutes an
ancilla qubit not subject to the symmetry action.

\subsection{Symmetries
\label{sec:discussion-sym}}

\subsubsection{Definition with restricted symmetry representations}

When defining phases under symmetries, we have allowed for arbitrary
representations of the symmetry group along the path. What if we want to
restrict to only the representation of the initial and final symmetry,
$U_g^0$ and $U_g^1$, respectively? It turns out that does not pose a
restriction for compact groups, as long as at least one of the effective
representations $U_g^0$ or $U_g^1$ after blocking to the normal form is
faithful.  Namely, given such a faithful representation $U_g$, we have
that $\chi_U(g)=\tr\,U_g = |\tr[V_g]|^2\ge0$ (with $V_g\otimes\bar V_g$
the virtual realization of $U_g$), which implies that any representation
$W_g$ is contained as a subrepresentation in $U_g^{\otimes N}$ for $N$
large enough. (For finite groups, this follows as the multiplicity
$\tfrac{1}{|G|}\sum_g \chi_U(g)^N\bar\chi_W(g)$ is dominated by
$\chi_W(1)$ since $|\chi_U(g)|$ is maximal for $g=1$,
cf.~Ref.~\onlinecite{fulton-harris}; for Lie groups, this argument needs
to be combined with a continuity argument,
cf.~Ref.~\onlinecite{mathoverflow:58633}). Thus, starting from the
symmetry representation $U_g^0$ or $U_g^1$, we can effectively obtain any
representation needed for the interpolating path by blocking $N$ sites,
proving equivalence of the two definitions.

\subsubsection{Definition with only one symmetry representation}

If the two systems to be compared are invariant under the same symmetry
$U_g^0=U_g^1$, we might want to build a path invariant under that
symmetry, instead of considering the symmetry $U_g=U_g^0\oplus U_g^1$ as
in our definition. In fact, these two definitions turn out the be
equivalent, since we can easily map a path with symmetry $U_g$ to one with
symmetry $U_g^0=U_g^1$, and vice versa.  Given a path
$\ket{\psi(\lambda)}$ (with corrsponding Hamiltonian) with symmetry
$U_g^0$, we can add an unconstrained ancilla qubit
(cf.~Sec.~\ref{app:phasedef-ancillas}) at each site and consider the path
$\ket{\psi(\lambda)} \otimes(\sqrt{1-\lambda}\,\ket0+
\sqrt{\lambda}\,\ket{1})^{\otimes
N}$, thus embedding the system in a space with symmetry $U_g$.

Conversely, we can map any path with symmetry $U_g$ to a path with
symmetry $U_g^0=U_g^1$. To this end, we can use $U_g = U_g^0\otimes
\openone_2$ to interpret the path as a path involving one system with
$U_g$--symmetry and one unconstrained ancilla qubit per site, which again
can be understood as being part of a $(U_g^0)^{\otimes 2}$--invariant
subspace, cf.~Sec.~\ref{app:phasedef-ancillas}. Note that the
factorization $U_g=U_g^0\otimes \openone_2$ requires
$U_g^0$ and $U_g^1$ to have the same phase, which motivates why we chose
to lock the phase between $U_g^0$ and $U_g^1$ in such a scenario.

\subsubsection{Definitions with different classifications}

Defining phases in the presence of symmetries is suble, as different
definitions can yield very different classifications. In the following, we
discuss some alternative definitions and their consequences.

One possibility would be not to allow for an arbitrary gauge of the
phases of the symmetries in $U_g^0$ and $U_g^1$, but to keep the gauge
fixed.  In that case, the 1D representations in
(\ref{eq:app-inj-sform:sform}) and (\ref{eq:sym:noninj:2223}) can enter
the classification: While finite representations can be still removed by
blocking sites, continuous representations remain different even after
blocking and cannot be changed into each other continuously, and thus,
phases are additionally labelled by the continuous 1D representations of
the symmetry. 

This classification changes once more if one allows
to compare blocks of different lengths for the two systems: E.g., for
$\mathrm{U}(1)$ any two continuous representations $e^{in_0\vartheta}$ and
$e^{in_1\vartheta}$ of $\vartheta\in \mathrm{U}(1)$ can be made equal by
blocking $|n_1|$ and $|n_0|$ sites, respectively, as long as the signs of
$n_0$ and $n_1$ are equal. Therefore, in that case the phases are
additionally labelled by the signs of the 1D representations; this
case has been considered in Ref.~\onlinecite{chen:1d-phases-rg}.

To give an example where one might want to fix the phase relation between
$U_g^0$ and $U_g^1$, consider the two product states $\ket{0000\cdots}$ and
$\ket{0101\cdots}$ under $U(1)$ symmetry: Both $U_g^0$ and $U_g^1$ arise
as subblocks of $R_\phi\otimes R_\phi$ [with $R_\phi=\exp(i
\phi Z/2)$ the original $\mathrm U(1)$ representation], which suggests to
fix the phase relation and in turn separates the two phases; note that
this can be seen as a way to reinforce translational invariance.

\subsubsection{A definition with only one phase}

We are now going to present an alternative definition of phases under
symmetries which yields a significantly different classification, namely
that all systems with the same ground state degeneracy are in the same
phase, just as without symmetries. The aim of this discussion is to point
out that it is important to fix the representation of the symmetry group,
and not only the symmetry group itself, in order to obtain a meaningful
classification of phases in the presence of symmetries.

We impose that along the path, the system is invariant under some (say,
faithful) representation of the symmetry group, which may however change
along the path. Then, however, it is possible to transform any state to
pairs of maximally entangled states between adjacent sites in the
\emph{same} basis, and therefore to the same parent Hamiltonian, by
rotating the isometric form. Thus, all injective systems can be
transformed to the same Hamiltonian preserving symmetries, and similarly
for symmetry broken systems. It follows that even with symmetries, all
systems are in the same phase as long as they have the same ground state
degeneracy. 

Note that while in our approach, we also use rotations to bring the system
into a simple form, these ``rotations'' should just be thought of as
choosing a convenient basis, and not as actual rotations. In
particular, due to our way of imposing the two symmetry representations on
orthogonal subspaces, $U_g=U^1_g\oplus U^2_g$, we never need to fix two
different bases for the same subspace.

\subsubsection{Projective symmetries }

While our definition of phases under symmetries can be applied to both 
linear and projective representations, we found that in the injective
case, symmetries are in fact always linear
(cf.~Appendix~\ref{app:stform-inj-sym}).   In the non-injective case,
however, there exist systems having projective symmetries, such as the
Majumdar-Ghosh model, which has the ground states
$(\ket{01}-\ket{10})^{\otimes N}$ and the same state 
translated by one lattice site.
This model is invariant under $\mathrm{SU}(2)/\mathbb Z_2$ [which can be
understood as a projective representation of $\mathrm{SO}(3)$] and under
the Pauli
matrices (a
projective representation of $D_2=\mathbb Z_2\times \mathbb Z_2$).  On the
other hand, any $d$-dimensional projective representation becomes linear
(up to trivial phases) after taking its $d$'th power; thus, 
classifying phases under linear symmetries as we did is in fact sufficient. 
Alternatively, any projective representation can be lifted to a linear
representation, which is still a symmetry of the Hamiltonian: E.g., any
$U_g$--invariant Hamiltonian, $[H,U_g^{\otimes N}]=0$, is also invariant
under $e^{i\phi}U_g$, and thus under the representation $V_k\equiv k$ of the
group $K=\langle U_g\rangle$ generated by the $U_g$ by itself.

On the other hand, in case we want to build a joint symmetry
representation $U_g=U_g^0\oplus U_g^1$ \emph{before} blocking, and do not
want to lift the joint representation to a larger group, we get
constraints on $U_g^0$ and $U_g^1$: Firstly, both of them need to be in
the same cohomology class---while the cohomology class can be changed by
blocking, one might want to compare systems with a particular notion of
locality. Also note that even if both symmetries are in the same
cohomology class, one needs to adjust the trivial phases such that
$\omega(g,h)$ is actually equal up to a one-dimensional representation of
the group, since otherwise $U_g$ does not form a representation;  note
however that this is actually a consequence of our requirement that $U_g$
forms a representation and does not follows from an underlying symmetry.

\subsubsection{Multiple copies and phases as a resource}

An interesting observation is that the classification of 1D phases under
symmetries is not stable if one takes multiple copies. For instance, one
can construct a path of smooth gapped Hamiltonians which interpolates from
two copies of the AKLT state to the trivial state while preserving
$\mathrm{SO}(3)$ symmetry. More generally, for any two states
$\MPS{\mc P_0}$ and $\MPS{\mc P_1}$ there exist $k_0$ and $k_1$ such that
$\MPS{\mc P_0}^{\otimes k_0}$ can be converted to $\MPS{\mc P_1}^{\otimes
k_1}$. This follows from the observation made in Sec.~\ref{sec:2d-sym} for
two-dimensional systems: Taking tensor products changes the projective
representation on the bond, and it is always possible to obtain 
a linear representation by taking a finite number of copies.

From a quantum information perspective, this shows that MPS which belong
to different phases should not be regarded as a resource such as
entanglement, but rather as characterized by conserved quantities such as
parity (i.e., described by a finite group). The minimum requirement
for a resource should be that it cannot be created ``for free'' (e.g., by
the quasi-local evolution created by a gapped
path~\cite{hastings:quasi-adiabatic}). However,  an arbitrary
even number of copies of the AKLT state can be created from one trivial
state, which demonstrates that phases under symmetries should not be
considered resources.

\subsubsection{Other symmetries}

While we have discussed the classification of phases for local symmetries
$U_g^{\otimes N}$, very similar ideas can be used to classify phases under
global symmetries such as inversion or time reversal
symmetry.~\cite{pollmann:symprot-1d,pollmann:1d-sym-protection-prb,chen:1d-phases-rg}
The fundamental concept---that two $\mc P$ representing the
same MPS or PEPS are related by a gauge transformation---equally applies
in the case of global symmetries. However, it should be noted that there
is an essential difference, in that the representation structure of the
global symmetry need not lead to a representation structure on the virtual
level, which in turn leads to classification criteria beyond cohomology classes.
Let us illustrate this for reflection symmetry: Reflection is realized by
applying a flip (swap) operator $\flip$ to the \emph{virtual} system,
together with an operation $\pi$ on the physical system reversing the
ordering of the blocked sites.  Thus, for an injective MPS $\MPS{\mc P}$
with reflection symmetry, we have that
\[
\pi\, \mc P\, \flip = \mc P (V_{-1}\otimes \bar V_{-1})\ ,
\]
where $V_{-1}$ is the virtual representation of the non-trivial element of
$\mathbb Z_2\equiv\{+1,-1\}$. 
(Note that if $\mc P$ is injective, $\pi$ cannot be trivial, since
otherwise $\flip=V_{-1}\otimes \bar V_{-1}$ which is impossible; this shows that an
injective MPS cannot have reflection symmetry unless it contains more than
one site per block.)
Applying a second reflection, we find that
\begin{linenomath}
\begin{align*}
\mc P &= \pi(\pi\mc P\, \flip)\,\flip 
= \pi\mc P\, (V_{-1}\otimes \bar V_{-1})\,\flip\\
&= (\pi\mc P\,\flip) \,(\bar V_{-1}\otimes V_{-1})
= \mc P\, (V_{-1}\bar V_{-1}\otimes \bar V_{-1}V_{-1})
\end{align*}
\end{linenomath}
i.e., the $\mathbb Z_2$ group structure of the symmetry is represented on
the virtual level as $V_{-1}\bar V_{-1} = e^{i\phi} \openone$---similar to a
projective representation, but with an additional complex conjugation. This
relation allows for phases $e^{i\phi}=\pm 1$, corresponding to symmetric
and antisymmetric unitaries $V_{-1}$, which cannot be connected continuously and thus
label different phases; this observation has been used in
Refs.~\onlinecite{pollmann:symprot-1d,pollmann:1d-sym-protection-prb}
to prove the separation of the AKLT phase from the trivial phase under either time
reversal or inversion symmetry.

\subsection{Examples
\label{sec:examples}}

\subsubsection{The six phases with $D_2$ symmetry}

As an example for the classification of one-dimensional phases in the
presence of symmetries, let us discuss the different phases under 
$D_2=\mathbb Z_2\times \mathbb Z_2$ symmetry, which appears e.g.\ as a
subsymmetry of $\mathrm{SO}(3)$ invariant models; we will see that there
is a total of six different phases under $D_2$ symmetry.

Let us label the elements of $D_2=\mathbb Z_2\times \mathbb Z_2$ by
$e\equiv(0,0)$, $x\equiv(1,0)$, $z\equiv(0,1)$, and $y\equiv(1,1)$, with
componentwise addition modulo $2$. $D_2=\{e,x,y,z\}$ forms a subgroup of
$\mathrm{SO}(3)$ by identifying $(1,0)$ with an $x$--rotation by $\pi$ and
$(0,1)$ with a $z$--rotation by $\pi$.  There are two equivalence classes
of projective representations, corresponding to the integer and
half-integer representations of $\mathrm{SO}(3)$. In particular, the
one-dimensional spin-$0$ representation
$\rho^1_e=\rho^1_x=\rho^1_y=\rho^1_z=1$ belongs to the trivial class, and
the two-dimensional spin-$\tfrac12$ representation $\rho^2_x=X$,
$\rho^2_y=Y$, and $\rho^2_z=Z$ (with $X$, $Y$, $Z$ the Pauli matrices)
belongs to the non-trivial class. For the following examples, we will
always consider systems with physical spin $S=1$; we label the basis
elements by their $S_z$ spin component, $\ket{\!-\!1}$, $\ket{0}$, and
$\ket{1}$,  and denote the physical representation of the symmetry group by 
\begin{linenomath}
\begin{align*}
    R_x=\exp[i\pi S_x]=&
        -\ket{-1}\bra{+1}-\ket{0}\bra{0}-\ket{+1}\bra{-1}\ ,\\
    R_z=\exp[i\pi S_z]=& 
        -\ket{-1}\bra{-1}+\ket{0}\bra{0}-\ket{+1}\bra{+1}\ ,\\
    R_y=\exp[i\pi S_y]=&
        +\ket{-1}\bra{+1}-\ket{0}\bra{0}+\ket{+1}\bra{-1} \ .
\end{align*}
\end{linenomath}

For systems with unique ground states, there are two possible phases.
One contains the trivial state
$\ket{0,\dots ,0}$,
which can be trivially written as an MPS with bond dimension $D=1$, and
\[\mc P_\mathrm{triv} = \ket{0}\ ;
\]
clearly, applying any physical transformation $R_w$ ($w=x,y,z$) to the
physical spin translates to applying the one-dimensional
representation $1\otimes 1$ on the virtual system. The corresponding
Hamiltonian is 
\begin{equation}
        \label{eq:example-D2-triv-parent}
H_\mathrm{triv} = -\sum_i \ket{0}\bra{0} + \mathrm{const}\ .
\end{equation}
The second phase with unique ground state is illustrated by the
AKLT state,~\cite{aklt} which can be written as an MPS with $D=2$, and a
projector
\[
\mc P_\mathrm{AKLT} = \Pi_{S=1} (\openone \otimes iY)\ ,
\]
with $\Pi_{S=1}$ the projector onto the $S=1$ subspace, and $iY=
\left(\begin{smallmatrix}0&1\\-1&0\end{smallmatrix}\right)$. Since we
have that $R_x = \Pi_{S=1} X\otimes X\Pi_{S=1}$, and correspondingly for
$y$ and $z$, it follows that the symmetry operations are represented on
the virtual level as $R_x \mc P = \mc P(X\otimes \bar X)$, $R_y \mc P =
\mc P(Y\otimes \bar Y)$,  and $R_z \mc P = \mc P(Z\otimes \bar Z)$, which
is a non-trivial projective representation of $\mathrm{SO(3)}$.   The
corresponding Hamiltonian is the AKLT Hamiltonian
\[
H_{\mathrm{AKLT}} = \sum_i \big[\vec S_i\cdot \vec S_{i+1} +
\tfrac13 (\vec S_i\cdot \vec S_{i+1})^2+\mathrm{const}\big]\ .
\]

In order to classify all phases with symmetry breaking, we need to
consider all proper subgroups of $D_2$. There are four of them:
\begin{linenomath}
\begin{align*}
H_x &=\{e,x\} \\
H_z &=\{e,z\} \\
H_y &=\{e,y\} \mbox{\ and}\\
H_{\mathrm{triv}} &= \{e\}
\end{align*}
\end{linenomath}
The first three are isomorphic to $\mathbb Z_2$, which has only trivial
projective representations, and thus, each of them labels one phase; since
$H_\mathrm{triv}$ also has only trivial representations, it corresponds to
a fourth phase with symmetry breaking.  The number of
symmetry broken ground states is $|D_2/H|$, i.e., the first three
cases have two-fold degenerate ground states, and the last case a
four-fold degenerate one.

Let us start with $H_z$. A representant of that phase is a GHZ-type state
of the form 
\[
\ket{\mathrm{GHZ}_z} = \ket{+1,\dots,+1} + \ket{-1,\dots,-1}\ ,
\]
 which can be written as an MPS with $D=2$ and
\[
\mc P_{\mathrm{GHZ}_z} = \ket{+1} \bra{0,0} + \ket{-1}\bra{1,1}\ ,
\]
where the basis elements $\ket{0}$ and $\ket{1}$ correspond to two
symmetry broken sectors.  The action of the symmetry on the virtual level
is 
\begin{linenomath}
\begin{align*}
R_x\mc P_{\mathrm{GHZ}_z} &= 
-\mc P_{\mathrm{GHZ}_z}\left[ \ket{0,0}\bra{1,1}+\ket{1,1}\bra{0,0}\right]\\ 
R_z\mc P_{\mathrm{GHZ}_z} &= 
-\mc P_{\mathrm{GHZ}_z}\left[ \ket{0,0}\bra{0,0}+\ket{1,1}\bra{1,1}\right]
\end{align*}
\end{linenomath}
and correspondingly for $R_y$; i.e., while $R_z$ acts within the symmetry
broken sectors, $R_x$ (and $R_y$) acts by permuting the different sectors.
The corresponding Hamiltonian is the GHZ Hamiltonian
\[
H_{\mathrm{GHZ}_z} =\! -\!\sum_i \big[
    \ket{+1,+1}\bra{+1,+1}+ \ket{-1,-1}\bra{-1,-1}
    \big]_{i,i+1} \ .
\]
The same type of ground state and Hamiltonian is found for the subgroups
$H_x$ and $H_y$ with correspondingly
interchanged roles (i.e., $R_x$ and $R_y$, respectively, do not permute
the ground states).  Note that these three phases are indeed distinct in
the presence of $D_2$ symmetry, as there is no way how to smoothly change
the element of the symmetry group which does not permute the symmetry
broken sectors. (This is related to the fact that $D_2/H$ is discrete,
i.e., we are breaking a discrete symmetry; note that breaking of
continuous symmetries does not fit the MPS framework since this would
correspond to an infinite number of blocks in the MPS and would
require gapless Hamiltonians.)

Finally, choosing $H_\mathrm{triv}=\{e\}$ gives a phase which fully breaks
the $D_2$ symmetry. A representative of this phase can be constructed by 
blocking two sites, with the four basis states
\begin{linenomath}
\begin{align*}
\ket{\hat 1} = \ket{+1}\ket{+x}\ ,
\quad&\ket{\hat 2} = \ket{+1}\ket{-x}\ ,\\
\ket{\hat 3} = \ket{-1}\ket{+x}\ ,
\quad&\ket{\hat 4} = \ket{-1}\ket{-x}\ ,
\end{align*}
\end{linenomath}
where $\ket{\pm x}$ are the $S_x$ eigenstates with eigenvalues $\pm 1$,
\[
\ket{\pm x} \propto \ket{-1} \pm \sqrt{2}\ket{0} + \ket{+1}\ .
\]
We have that 
\begin{linenomath}
\begin{align*}
R_x\ket{\pm x} &= -\ket{\pm x}\ ,& R_z\ket{\pm 1} &= -\ket{\pm 1}\ ,\\
R_x\ket{\pm 1} &= -\ket{\mp 1}\ ,& R_z\ket{\pm x} &= -\ket{\mp x}\ .
\end{align*}
\end{linenomath}
It follows that all $R_w\otimes R_w$ ($w=x,y,z$) act as permutations on
the basis $\{\ket{\hat 1},\ket{\hat 2},\ket{\hat 3},\ket{\hat 4}\}$: 
\begin{linenomath}
\begin{align*}
(R_x\otimes R_x):\  &\ket{\hat 1}\leftrightarrow \ket{\hat 3}\,; \ 
     \ket{\hat 2}\leftrightarrow \ket{\hat 4}\,; \\
(R_z\otimes R_z):\  &\ket{\hat 1}\leftrightarrow \ket{\hat 2}\,; \ 
     \ket{\hat 3}\leftrightarrow \ket{\hat 4}\,; \\
(R_y\otimes R_y):\  &\ket{\hat 1}\leftrightarrow \ket{\hat 4}\,; \ 
     \ket{\hat 2}\leftrightarrow \ket{\hat 3}\,.
\end{align*}
\end{linenomath}
Thus, the GHZ type state 
\[
\ket{\mathrm{GHZ}_4} = \sum_{k=1}^{4} \ket{\hat k,\dots, \hat k}\ ,
\]
which is an MPS with $D=4$ and 
\[
\mc P_{\mathrm{GHZ}_4} = \sum_{k=1}^4\ket{\hat k}\bra{k,k}\ ,
\]
breaks all symmetries of $D_2$, and represents yet another distinct phase
with symmetry $D_2=\mathbb Z_2\times \mathbb Z_2$; the corresponding
Hamiltonian is 
\[
H_{\mathrm{GHZ}_4} = 
-\sum_j \Big[\sum_{k=1}^{4} \ket{\hat k,\hat k}\bra{\hat k,\hat k}_{j,j+1}\Big]\ ,
\]
where the label $j$ refers to the blocked sites.

\subsubsection{Phases under $\mathrm{SO}(3)$ and
$\mathrm{SU}(2)$ symmetry}

Let us now consider symmetry under rotational invariance, imposed either
as $\mathrm{SO}(3)$ or as $\mathrm{SU(2)}$ symmetry. We will find that
under $\mathrm{SO}(3)$ symmetry, there are two possible phases,
represented by the spin-$1$ AKLT state and the trivial spin-$0$ state,
respectively; on the other hand, we will show that if we impose
$\mathrm{SU}(2)$ symmetry, there is only a single phase, as the AKLT state
can be transformed into the trivial state keeping $\mathrm{SU}(2)$
symmetry.

Let us start with $\mathrm{SO}(3)$ symmetry. In order to compare the
AKLT state to the trivial spin-$0$ state, we need a representation of
$\mathrm{SO}(3)$ which contains both the spin-$1$ and spin-$0$
representation; we will denote the representation as
\[
R_{\hat{\bm n}}(\theta) = \exp[i\, \theta\, \hat{\bm n}\cdot \bm S]\oplus 1\ .
\]
While we could start with such a symmetry representation right away, let
us discuss how to obtain the same setting from a spin-$1$ chain by
blocking: Blocking two
spin-$1$ sites gives a system with total spin $1\otimes 1=2\oplus 1 \oplus
0$, containing both a spin-$1$ and spin-$0$ subspace. It is
straightforward to check that after blocking two sites and applying a
rotation $1\otimes iY$, the isometric form of the AKLT state is
\begin{equation}
        \label{eq:ex:aklt-iso-proj}
        \hat{\mc P}_{\mathrm{AKLT}} = \openone\ ,
\end{equation}
where the identity is on the two virtual spins with representation
$\tfrac12\otimes \tfrac12 = 1\oplus 0$.  The physical 
rotation $R_{\hat{\bm n}}$
acts on the virtual indices with the 
projective spin-$\tfrac12$ representation of the rotation group; it
follows that the AKLT state is in the non-trivial equivalence class under
$\mathrm{SO}(3)$ symmetry.  The trivial spin-$1$ state, on the other hand,
is obtained by placing singlets between pairs of spin-$1$ sites [i.e.,
between sites
$(1,2)$, $(3,4)$, etc.]. After blocking these pairs, we obtain a product
state with the spin-$0$ state at each site, which is an MPS with $D=1$
and the trivial projector 
\begin{equation}
        \label{eq:ex:trivialstate-iso-proj}
        \hat{\mc P}_{\mathrm{triv}} = \ket{S=0}\ .
\end{equation}
Thus, the rotation group $R_{\hat{\bm n}}(\theta)$ acts with the trivial
representation on the virtual indices, and the trivial state is thus in a
different phase than the AKLT chain.

It should be noted that while $D_2$ is a subgroup of
$\mathrm{SO}(3)$, this does not imply that $\mathrm{SO}(3)$ exhibits all
phases of $D_2$---indeed, the symmetry broken phases are missing. The
reason is that while for any subgroup $H\subset D_2$, $D_2/H$ is finite,
this is not true for $\mathrm{SO}(3)$. Since however
$\mathrm{SO}(3)/H$ labels the symmetry broken ground states, this
corresponds to breaking a continuous symmetry, which leads to gapless
phases and cannot be described in the framework of Matrix Product States.

Let us now turn our attention towards $\mathrm{SU}(2)$ symmetry, and
explicitly construct an interpolating path between the isometric
projectors for the AKLT state, Eq.~(\ref{eq:ex:aklt-iso-proj}), and the
trivial state, Eq.~(\ref{eq:ex:trivialstate-iso-proj}). Note that the
difference is that now for both $\hat P_\mathrm{AKLT}$ and $\hat
P_\mathrm{triv}$, the symmetry action on the virtual level is a linear
representation of $\mathrm{SU}(2)$ (namely the spin-$\tfrac12$ and the
spin-$0$ representation, respectively). We can now provide an
interpolating path $\ket{\Psi_\gamma}=\ket{\omega(\gamma)}^{\otimes N}$,
with 
\[
\ket{\omega(\gamma)} = \gamma \ket{0,0} + (1-\gamma)
(\ket{1,1}+\ket{2,2})\ ,
\]
where $\ket{\Psi_0}$ corresponds to the isometric form
(\ref{eq:ex:aklt-iso-proj}) of the AKLT state, and $\ket{\Psi_1}$ to the
isometric from (\ref{eq:ex:trivialstate-iso-proj}) of the trivial state.
Furthermore, for $U\in \mathrm{SU}(2)$, the whole path is invariant under
the on-site symmetry 
\begin{equation}
        \label{eq:ex:su2-interpol-symmetry}
(1\oplus U)\otimes (1\oplus \bar U)
= 1\oplus (U\oplus \bar U)\oplus (U\otimes \bar U)
\end{equation}
which contains the symmetries $1$ and $U\otimes \bar U$ of the trivial and
the AKLT state as subsymmetries; this proves that under
$\mathrm{SU}(2)$ symmetry, the AKLT state and the trivial state are in
the same phase.  Note that the symmetry
(\ref{eq:ex:su2-interpol-symmetry}) is only a representation of
$\mathrm{SU}(2)$, but not of $\mathrm{SO(3)}$, as integer and half-integer
spin representations belong to inequivalent classes of projective
representations of $\mathrm{SO}(3)$; also, we cannot obtain this
symmetry by starting only from the spin-$0$ and spin-$1$ representation of
$\mathrm{SU}(2)$, as they are not faithful representations.
  Note that this interpolating
path can already be ruled out by imposing a parity constraint on the total
number of half-integer representations, by e.g.\ associating them to
fermions.

\section{Conclusion and Outlook}

In this paper, we have classified the possible phases of one-dimensional,
and to a certain extent two-dimensional, systems in the framework of Matrix
Product States and PEPS. We have done so by studying Hamiltonians with exact
MPS and PEPS ground states, and classifying under which conditions it is
possible or impossible to connect two such Hamiltonians along a smooth and
gapped path of local Hamiltonians.

We have found that in the absence of symmetries, all systems are in the
same phase, up to accidental ground state degeneracies.  Imposing local
symmetries leads to a more refined classification: For systems with unique
ground
states, different phases are labelled by equivalence classes of projective
representations, this is, cohomology classes of the group; for systems
with degenerate ground states, we found that the symmetry action can be
understood as composed of a permutation (permuting the symmetry broken
ground states) and a representation of a subgroup (acting on the
individual ground states), which together form an induced representation,
and different phases are labelled by the permutation action (this is, the
subgroup) and the cohomology classes of the subgroup.  In this
classification, systems in the same phase can be connected along a path
which is gapped even in the thermodynamic limit, while for systems in
different phases, the gap along any interpolating path will close
even for a finite chain.

We have subsequently studied two-dimensional systems and considered three
classes of phases, namely product states, GHZ states, and topological
models based on quantum doubles. We have shown that all of these phases
are stable in some region, and demonstrated that within that region,
and more generally within the framework used for MPS, imposing
symmetries does not further constrain the phase diagram.

We have also compared different definitions of phases under symmetries and
found that very different classifications can be obtained depending on the
definition chosen, ranging from scenarios where symmetries don't affect
the classification at all, to scenarios where the classification is
more fine-grained and e.g.\ the one-dimensional representations of the
group partly or fully enter the classification.  In this context, it is
interesting to note that there is a hierarchy in the classification of
phases as the spatial
dimension increases: Zero-dimensional phases are labelled by 1D
representations of the symmetry group (this is, its first cohomology
group).  This label vanishes in one dimension, and phases are now
classified by the second cohomology group. This label, in turn, vanishes
in three dimensions, and although we have demonstrated that we can't infer
symmetry constraints from the continuity of the PEPS projectors $\mc P$
alone, it is expected that phases under symmetries in two and more
dimensions are still classified by higher order cohomology
groups.~\cite{dijkgraaf:phases-cohomologies,kitaev:private}

A central tool in our proofs has been the \emph{isometric form} of an MPS
or PEPS. Isometric MPS and PEPS are fixed points of renormalization
transformations, and any MPS can be transformed into its isometric form
along a gapped path in Hamiltonian space; this result allows us to
restrict our classification of one-dimensional quantum phases to the case
of isometric RG fixed points.  Moreover, it gives us a tool to carry out
renormalization transformations in a local fashion, this is, without
actually having to block and renormalize the system; it thus provides a
rigorous justification for the application of RG flows towards the classification
of quantum phases. Let us add that the possibility to define an isometric
form, as well as the possibility to interpolate towards it along a
continuous path of parent Hamiltonians, still holds for not translational
invariant systems; however, without translational invariance we are
lacking tools to assess the gappedness of the Hamiltonian.

Let us note that MPS have been previously applied to the classification of
phases of one-dimensional quantum
systems:\cite{pollmann:symprot-1d,fidkowski:1d-fermions,chen:1d-phases-rg}
In particular, in Ref.~\onlinecite{pollmann:symprot-1d}, MPS have been
used to demonstrate the symmetry protection of the AKLT phase, and in
Ref.~\onlinecite{chen:1d-phases-rg}, renormalization
transformations~\cite{verstraete:renorm-MPS} and their fixed points on MPS
have been applied to the classification of quantum phases for one
dimensional systems with unique ground states both with and without
symmetries, giving a classification based on cohomology classes and 1D
representations.  Beyond that, RG fixed points of PEPS have also been used
towards the  classification of phases for two-dimensional
systems.\cite{chen:2d-rg,gu:fermion-phases-rg}

\subsection*{Acknowledgements}

We acknowledge helpful discussions with 
Salman Beigi, 
Oliver Buerschaper,
Steve Flammia,
Stephen Jordan, 
Alexei Kitaev,
Robert K\"onig, 
Spiros Michalakis,
John Preskill, 
Volkher Scholz,
Frank Verstraete,
and 
Michael Wolf.
This work has been supported by the Gordon and Betty Moore Foundation
through Caltech's Center for the Physics of Information, the NSF Grant 
No.~\mbox{PHY-0803371}, the ARO Grant No.~W911NF-09-1-0442,
the Spanish grants I-MATH, MTM2008-01366, and
\mbox{S2009/ESP-1594}, 
the European project QUEVADIS, and the DFG (Forschergruppe 635).

\appendix

\section{Gap proof for the 1D path}

In the following, we show that the family of $\gamma$-deformed parent
Hamiltonians which arise from the MPS path $\MPS{\mc P_\gamma}$ interpolating
between an MPS and its isometric form is gapped. Recall that this family
was defined as $H_\gamma:=\sum h_\gamma(i,i+1)$, with
$h_\gamma:=\Lambda_\gamma h_0\Lambda_\gamma>0$.  

We want to show that the path $H_\gamma$ is uniformly gapped, i.e., there
is a $\Delta>0$ which lower bounds the gap of $H_\gamma$ uniformly in
$\gamma$ and the systems size $N$: This establishes that the
$\MPS{\mc P_\gamma}$, and the corresponding $H_\gamma$,  are all in the same
phase. To this end, we use a result of
Nachtergaele~\cite{nachtergaele:degen-mps} (extending the result
of Ref.~\onlinecite{fannes:FCS} for the injective case), where it is shown that any
parent Hamiltonian is gapped, and a lower bound on the gap (uniform in
$N$) is given. 

In the following, we will use the results of 
Ref.~\onlinecite{nachtergaele:degen-mps}
to derive a uniform lower bound on the gap for all $H_\gamma$,
$1\ge\gamma\ge0$. Let the MPS matrices $[A_i(\gamma)]_{kl} 
:= \sum_{k,l}
\bra{i}\mc P_\gamma\ket{k,l} \ket{k}\!\bra{l}$
(cf.~Sec.~\ref{sec:disc:mps-with-matrices});
in the normal form, the $A_i(\gamma)$ have a block structure
$A_i(\gamma)=\bigoplus A_i^\alpha(\gamma)$.
Let $\mathbb E^\alpha(\gamma) := \sum_i A_i^\alpha(\gamma)\otimes
\overline{A_i^\alpha(\gamma)}$, and let $\left|\mathrm{spec}\,\mathbb
E^\alpha(\gamma)\right|=
\{\lambda_1^\alpha(\gamma)>\lambda_2^\alpha(\gamma)>\dots\ge0\}$ be the
ordered absolute value of the spectrum of $\mathbb E^\alpha(\gamma)$ (not
counting duplicates).  Then, $\lambda_2(\gamma)/\lambda_1(\gamma)<1$, and
since the spectrum is continuous in $\gamma\in[0;1]$, and the degeneracy
of $\lambda_1$ is $\mathcal A$,\cite{nachtergaele:degen-mps} the
existence of a uniform upper bound
$1>\tau_\alpha>\lambda_2^\alpha(\gamma)/\lambda_1^\alpha(\gamma)$ follows.
For $\alpha\ne\beta$, let
\[
\Omega_{\alpha,\beta}^p(\gamma)=\sup_{X,Y}
\frac{\big\langle\Phi[A^\alpha(\gamma);X]\big\vert\Phi[A^\beta(\gamma);Y]\big\rangle}{
    \big\|\ket{\Phi[A^\alpha(\gamma);X]}\big\|
    \big\|\ket{\Phi[A^\beta(\gamma);Y]}\big\|}\ ,
\]    
where $\ket{\Phi[C;X]}:= \sum_{i_1,\dots,i_p}\tr[C_{i_1}\dots
C_{i_p}X]\ket{i_1,\dots,i_p}$; i.e., $\Omega_{\alpha,\beta}^p(\gamma)$ is
the maximal overlap of the $p$-site reduced states of the MPS
described by the blocks $A^\alpha(\gamma)$ and $A^\beta(\gamma)$.
 With $\mathcal
S_\alpha(\gamma):=\{\sum_i\tr[A^\alpha_i(\gamma)X]\ket{i}|X\}$, 
and $\mathcal O(\mathcal X,\mathcal
Y)$ the maximal overlap between normalized vectors in the subspaces
$\mathcal X$ and $\mathcal Y$, we have that
$\Omega_{\alpha,\beta}^p(\gamma)
\le \mathcal O(\mathcal S_\alpha(\gamma)^{\otimes p},\mathcal
S_\beta(\gamma)^{\otimes p})
\le \mathcal O(\mathcal S_\alpha(\gamma),\mathcal S_\beta(\gamma))^p$.
Moreover, since $\mathcal S_\alpha(0)\perp\mathcal S_\beta(0)$, and
$\mathcal S_{\bullet}(\gamma)=Q_\gamma S_{\bullet}(0)$, we have that 
\begin{linenomath}
\begin{align*}
\mathcal O(\mathcal S_\alpha(\gamma),\mathcal S_\beta(\gamma))
&\le 
    \sup_{\bra{v}w\rangle=0}
    \frac{|\bra{v}Q_\gamma^2\ket{w}|}{
    \|Q_\gamma\ket{v}\|\|Q_\gamma\ket{w}\|}
\\
&=
    \sup_{\bra{v}w\rangle=0}
    \sqrt{\frac{|M_{12}|^2}{M_{11}M_{22}}}\ ,
\end{align*}
\end{linenomath}
where $M=\pi Q_\gamma^2\pi^\dagger$, 
$\pi=\ket{0}\!\bra{v}+\ket{1}\!\bra{w}$,
is some $2\times2$ submatrix of $Q_\gamma^2$. For $M>0$,
\begin{linenomath}
\begin{align*}
\frac{|M_{12}|^2}{M_{11}M_{22}}
   &\le
   1-\frac{\lambda_{\min}(M)}{\lambda_{\max}(M)}
 \le 
   1-\frac{\lambda_{\min}(Q_\gamma^2)}{\lambda_{\max}(Q_\gamma^2)}
\\
   &\le
   1-\lambda_{\min}(Q^2)=:\kappa
    < 1\ ,
\end{align*}
\end{linenomath}
and we find that $\Omega^p_{\alpha,\beta}(\gamma) \le \kappa^p$. 
Thus, there exists a $p$ s.th.\
\[
K^p(\gamma):=
\frac{4(\mathcal A-1)\kappa^p}{1-2(\mathcal A-1)\kappa^p}
+\sum_\alpha D^2 \tau^p_\alpha\frac{1+D^2\tau^p_\alpha}{1-D^2\tau^p_\alpha} 
<1/\sqrt{2}\ ,
\]
and as Nachtergaele shows,\cite{nachtergaele:degen-mps}
$\tfrac12\Delta_{2p}(\gamma)(1-\sqrt{2} K^p(\gamma))^2$ is a lower bound
on the spectral gap of $H_\gamma$. Here, $\Delta_{2p}(\gamma)$ is 
the gap of $H_\gamma$, restricted to $2p$ sites, which has a uniform
lower bound as the restricted Hamiltonian is continuous in $\gamma$.
This proves that $H_\gamma$ has a uniform spectral gap for $0\le\gamma\le
1$.

\section{Standard form for injective MPS under symmetries
\label{app:stform-inj-sym}}

In this section, we discuss how $U_g$--symmetry of an injective MPS is
represented on the virtual level.  To start with, it has been
shown~\cite{david:stringorder-1d} that any two tensors $\mc P$ and $\mc
P'$ which (up to a phase) represent the same MPS can be related by a gauge
transformation 
\[
\mc P' = \mc P (e^{i\phi} V\otimes \bar V)\ .
\]
Given an MPS $\MPS{P}$ with $U_g$--invariant parent Hamiltonian (where
$U_g$ is a linear or projective representation), it follows that $\MPS{\mc
P}$ is invariant under $U_g$ up to a phase and thus, the action of $U_g$
on $\MPS{\mc P}$ can be understood on the virtual level as
\begin{equation}
        \label{eq:app-inj-sform:sform}
U_g\mc P= \mc P (e^{i\phi_g} V_g\otimes \bar V_g)\ ,
\end{equation}
or
\[
\hat U_g :=\mc P^{-1} U_g\mc P= (e^{i\phi_g} V_g\otimes \bar V_g)\ ;
\]
note that $\hat U_g$ forms again a representation.  From this, it follows
that the $V_g$ form a projective representation, and in turn that $\phi_g$
forms a 1D representation of $G$. This shows that $\hat U_g$, and thus
$U_g$, is a \emph{linear} representation of $G$---systems with MPS ground
states without symmetry breaking cannot be invariant under projective
representations.  (Strictly speaking, we only find that $\phi_g$ is in the
same cohomology class as a linear representation, but in order to compare
two systems, we need the same gauge for $\phi_g^0$ and $\phi_g^1$, so we
choose them to be linear representations.)

Under blocking, the physical symmetry $U_g$ in (\ref{eq:app-inj-sform:sform})
is mapped to $U_g^{\otimes k}$, restricted to the
range of the blocked map $\mc P^{\otimes k}\ket{\omega_D}^{\otimes (k-1)}$. 
$V_g$ remains unchanged under blocking;
this suggests that it is suitable as a characteristic of a quantum
phase (which we expect to be stable under blocking).  The 1D
representation $e^{i\phi_g}$, on the other hand, changes under blocking to 
$e^{ik\phi_g}$. In particular, if the 1D representation is finite, it can
be removed by blocking, while this is not possible for continuous
representations such as of $\mathrm{U}(1)$. 

However, our definition of quantum phases allows us to fully remove the
1D representation $e^{i\phi_g}$ in (\ref{eq:app-inj-sform:sform}) even if
it is continuous, using
the phase degree of freedom which we included in our definition. Consider
first the scenario where we are free to choose the phase degree of freedom
for the initial and final system independently. Then, we can choose to
replace $U_g$ by $e^{i\phi_g}U_g$, which will make the phase in
(\ref{eq:app-inj-sform:sform}) vanish. Second, consider the case where the
two physical symmetries are equal, $U_g^0=U_g^1=:U_g$. Then, we can choose
the phase gauge such that the 1D representation for one system vanishes, 
\[
U_g^0=V^0_g\otimes \bar V_g^0\ .
\]
On the other hand, since $U_g^0=U_g^1$, we have that
\[
e^{i\phi_g^1}V^1_g\otimes \bar V_g^1
= V_g^0\otimes \bar V_g^0
\]
which also implies that $e^{i\phi_g^1}=1$ [e.g., by looking at any
non-zero matrix element $(i,i)\times(j,j)$]. Finally, if $U_g^0$ and
$U_g^1$ only share a subblock, this means that $V_g^0=X_g^0\oplus Y_g^0$
s.th.\ $X_g^0\otimes\bar X_g^0$ yields that subblock, and similary for
$V_g^1$, which again shows that $e^{i\phi_g^1}=1$.

\section{Continuity of cohomology class where subspace changes
    \label{app:coh-class-cont-dimchange}}

This appendix contains the proof omitted at the end of
Sec.~\ref{subsubsec:sym-inj-separation} (all the notation is the same as
introduced there): That the cohomology class obtained from $\mc P_\gamma$
and $\mc P_{\gamma+\dd\gamma}$ is the same even if $\mc
P_{\gamma+\dd\gamma}$ is supported on a larger space than $\mc P_\gamma$.
We will again show this for differentiable $\mc P_\gamma$, 
it extends to continuous $\mc P_\gamma$ by approximating them by a
sequence of differentiable functions.  Let
$\mc P_{\gamma+\dd\gamma}= \mc P_\gamma+\mc P'_\gamma\dd\gamma$, 
with
\[
\mc P_\gamma = \left[\begin{matrix}P&0\\0&0\end{matrix}\right]
\]
and
\[
\mc P'_\gamma = \left[\begin{matrix}A&B\\C&D\end{matrix}\right]\ ,
\]
and note that the same block structure has to hold for the physical
symmetry (as it is a symmetry for the whole path),
\[
U_g = \left[\begin{matrix}Q_g & 0 \\ 0 & R_g\end{matrix}\right]\ ,
\]
with $R_g$, $S_g$ unitary.  To first order in $\dd\gamma$, the action of
the symmetry on the virtual space is
\[
\tilde U_g = 
    \mc P_{\gamma+\dd\gamma}^{-1} U_g \mc P_{\gamma+\dd\gamma} 
= T^{-1} 
K_g
    T+O(\dd\gamma^2)\ ,
    \]
with
\begin{linenomath}
\begin{align*}
K_g&=\left[\begin{array}{cc}
    (\openone+X)\bar Q_g - X\bar R_g & \bar Q_g X - X \bar R_g \\[4pt]
    (\openone+X)(\bar R-\bar Q)  & (\openone+X)\bar R_g-\bar Q_g X
    \end{array}\right]\ ,
    \\[4pt]
T&=\left[\begin{matrix} \openone & 0 \\0&C^{-1}D\end{matrix}\right]\ ,
\end{align*}
\end{linenomath}
$X=(P+A\,\dd\gamma)^{-1}BD^{-1}C\dd\gamma$, and the ``dressed representations'' 
\begin{linenomath}
\begin{align*}
\bar Q_g &= (P+A\,\dd\gamma)^{-1} Q_g (P+A\,\dd\gamma)\ ,\\
\bar R_g &= C^{-1}R_g C\ .
\end{align*}
\end{linenomath}
(This can be derived using the Schur complement to express $\mc
P^{-1}_{\gamma+\dd\gamma}$ and can be readily checked by left
multiplication with $\mc P_{\gamma+\dd\gamma}$.) It follows that the upper
left block of $K_gK_h$ is 
\[
(\openone+X)\bar Q_g\bar Q_h - X\bar R_g\bar R_h + O(\dd\gamma^2)\ ,
\]
i.e., the upper left block of $U_g$ (which corresponds to the
virtual subspace used by $\mc P_\gamma$) forms a representation to first
order in $\dd\gamma$. The part corresponding to each of the two bonds thus
forms a projective representation which changes smoothly in $\gamma$, and
which therefore cannot change its cohomology class.  Since the cohomology
class has to be the same for all subblocks, this completes the proof.

\section{Standard form for non-injective MPS under symmetries
\label{app:stdform-noninj-sym}}

In the following, we will show that for any non-injective MPS $\MPS{\mc
P}$ with a parent Hamiltonian which is invariant under some local symmetry
$U_g$, the symmetry can be represented as the symmetry 
(\ref{eq:sym:noninj:generalform}) of $\mc P$.
It is clear from (\ref{eq:mps-def:parentham-def}) that any $\mc P$ with
this symmetry will result in a Hamiltonian with the same symmetry. In the
following, we will prove the converse.

Using the same arguments as in Ref.~\onlinecite{sanz:mps-syms}, one can
show that any $U_g$--invariance of a parent Hamiltonian (at the moment, 
we are talking of
a \emph{single} unitary $U_g$ and only carry the subscript $g$ for
``future use'') can be understood
as resulting from an effective action
\begin{equation}
\label{eq:sym:noninj:2223}
\hat U_g:=
\mc P^{-1}\, U_g\:\mc P = P_g\bigg[\bigoplus_{\alpha} e^{i\phi^\alpha_g}
    V_g^\alpha\otimes \bar V_g^\alpha\bigg]
\end{equation}
of the symmetry operation on the virtual system, using the correct gauge for
$\mc P$. Here, $V_g^\alpha$ is unitary, the direct sum runs over the
Hilbert spaces $\mc H_\alpha$ in (\ref{eq:noninj:diagonalspace}), and
$P_g$ permutes those Hilbert spaces.
(Proof sketch, cf.~Ref.~\onlinecite{sanz:mps-syms}: $U_g$--invariance of
the Hamiltonian means that $U_g$ maps ground states to ground states. The
different sectors $\mc H_\alpha$,
corresponding to different symmetry broken sectors, must be treated
independently, since they don't interfere on any OBC interval. Thus, $U_g$
can on the one hand permute sectors and change their phase---this gives
the $P_g$ and $\exp[i\phi^\alpha_g]$---and it can on the other hand act
nontrivially on each sector---since each of the sectors behaves like an
injective MPS, this gives the $V^\alpha_g\otimes \bar V_g^\alpha$.)

In the following, we will assume that $P_g$ acts irreducibly on the system
in the sense that there are no subsets of $\{1,\dots,\mc A\}$ invariant
under all $P_g$.  We can always achieve this situation by splitting
(\ref{eq:sym:noninj:2223})
into a direct sum over such irreducible cases.

Let us now study what a linear representation structure 
\begin{equation}
        \label{eq:app-noninj-sym:Ug-repstruct}
\hat U_g\hat U_h = \hat U_{gh}
\end{equation}
($g,h\in G$) implies for the algebraic structure of $P_g$, $V_g^\alpha$,
and $e^{i\phi_g^\alpha}$. (We can always achieve a linear representation
by blocking; also note that the following argument can be generalized to
projective representations.) First, since $P_g$ is the only part of
(\ref{eq:sym:noninj:2223}) which is not block diagonal, it follows that
the $P_g$ form a linear representation of $G$. (Linearity follows since
the entries of $P_g$ are $0$ and $1$). Let us define 
\begin{equation}
        \label{eq:app-noninj-sym:def-W}
W^\alpha_g:= e^{i\phi^\alpha_g} V_g^\alpha\otimes \bar V_g^\alpha\ .
\end{equation}
Then, using (\ref{eq:sym:noninj:2223}) the representation
structure (\ref{eq:app-noninj-sym:Ug-repstruct}) is equivalent to the
relation
\begin{equation}
        \label{eq:app-noninj-sum:Wg-structure}
W^{\pi_h(\alpha)}_g W^\alpha_h =W^\alpha_{gh}
\end{equation}
for the $W^{\alpha}_g$, where the permutations $\pi_h$ are defined via
$\mc H_{\pi_h(\alpha)}=P_h(\mc H_{\alpha})$ and thus form a representation,
$\pi_g\pi_h = \pi_{gh}$.

Let us now show that (\ref{eq:app-noninj-sum:Wg-structure}) implies that
the $\hat U_g$ can be understood as an induced representation; the following
proof is due to S.~Beigi.\cite{beigi:private} Fix some $\alpha_0$, and let
\[
H:=\{h:h\in G, \pi_h(\alpha_0)=\alpha_0\}\ .
\]
Then, (\ref{eq:app-noninj-sum:Wg-structure}) implies that $W_h^{\alpha_0}$
is a linear representation of $H$. We know we can write $G$ as the
disjoint union over cosets $k_\beta H$ labelled by the blocks
$\beta=1,\dots,\mc A$, for a (non-unique) choice of $k_\beta\in G$
chosen such that $\pi_{k_\beta}(\alpha_0)=\beta$. (This is where we need
irreducibility.)
We now have that
\begin{equation}
        \label{eq:app-noninj-sym:prove-ind-rep-1}
W_g^\beta 
\stackrel{(\ref{eq:app-noninj-sum:Wg-structure})}{=}
W_{gk_\beta}^{ \pi_{k_\beta^{-1}}\!(\beta)} W_{k_\beta^{-1}}^\beta
=
W_{gk_\beta}^{\alpha_0} W_{k_\beta^{-1}}^\beta 
\end{equation}
where we have used that
$\pi_{k_\beta^{-1}}(\beta)=\pi^{-1}_{k_\beta}(\beta)=\alpha_0$.
Using the decomposition of $G$ in cosets, we have that $g$ and $\beta$
uniquely determine $\gamma$ and $h\in H$ by virtue of
\begin{equation}
    \label{eq:app-noninj-sym:hgamma-from-gbeta}
gk_\beta = k_\gamma h
\end{equation}
and thus, continuing (\ref{eq:app-noninj-sym:prove-ind-rep-1}),
\[
W_g^\beta 
\stackrel{(\ref{eq:app-noninj-sum:Wg-structure})}{=}
W_{k_\gamma}^{\pi_h(\alpha_0)} 
W^{\alpha_0}_h W_{k_\beta^{-1}}^\beta
= 
 W_{k_\gamma}^{\alpha_0}
W^{\alpha_0}_h W_{k_\beta^{-1}}^\beta\ ,
\]
using that $\pi_h(\alpha_0)=\alpha_0$. We can now 
replace $W^\beta_{k_\beta^{-1}}$ using that
\[
\openone =
W_{k_\beta k_\beta^{-1}}^\beta 
\stackrel{{(\ref{eq:app-noninj-sum:Wg-structure})}}{=}
W^{\pi_{k_\beta^{-1}}(\beta)}_{k_\beta} W^\beta_{k_\beta^{-1}}
=
W^{\alpha_0}_{k_\beta} W^\beta_{k_\beta^{-1}}\ .
\]
Substituting this above, we finally obtain that
\begin{equation}
        \label{eq:app-noninj-sym:Wg-induced-twisted}
W_g^\beta = W_{k_\gamma}^{\alpha_0}\;
W^{\alpha_0}_h\: \big(W_{k_\beta}^{\alpha_0}\big)^{-1}\ ,
\end{equation}
i.e., $W_g^\beta$ is fully determined by the representation
$W_h^{\alpha_0}$ on $H$, together with the (arbitrary) unitaries
$W_{k_\beta}^{\alpha_0}$ for all coset representatives $k_\beta$.

We now define a rotation
\[
K=\bigoplus_\delta W_{k_\delta}^{\alpha_0}\ ;
\]
then, in the rotated basis, $\hat U_g$ reads
\[
K^\dagger \hat U_g K = 
P_g\bigg[\bigoplus_\beta  W^{\alpha_0}_h \bigg]\ ,
\]
using that 
\mbox{$\pi_g(\beta) = \pi_{k_\gamma}\big(\pi_h\big(
\pi_{k_\beta^{-1}}(\beta)\big)\big)
= \pi_{k_\gamma}\big(\pi_h(\alpha_0)\big)
=\gamma$}.
If we now substitute back
\[
W_h^{\alpha_0} = e^{i\phi^{\alpha_0}_h} V_h^{\alpha_0}\otimes \bar
V_h^{\alpha_0}
\]
[Eq.~(\ref{eq:app-noninj-sym:def-W})] in 
\[
W_g^{\alpha_0} W_h^{\alpha_0} =
W_{gh}^{\alpha_0}\ ,\quad g,h\in H
\]
[Eq.~(\ref{eq:app-noninj-sum:Wg-structure})], we find that 
$V_h^{\alpha_0}$ forms a
projective representation of $H$, and $e^{i\phi_h^{\alpha_0}}$ a linear
representation of $H$. 

Thus, we can write
\[
K^\dagger \hat U_g K = \tilde U_g D_g
\]
with 
\begin{equation}
        \label{eq:app-noninj-sym:final-form-irrep}
\tilde U_g = 
P_g\bigg[\bigoplus_\beta V^{\alpha_0}_h\otimes\bar V^{\alpha_0}_h \bigg]
\end{equation}
and
\begin{equation}
        \label{eq:app-noninj-sym:diagonal-part-irrep}
D_g = \bigoplus_\beta e^{i\phi_h^{\alpha_0}}\ ,
\end{equation}
where $h\equiv h(g,\beta)$ is determined by
(\ref{eq:app-noninj-sym:hgamma-from-gbeta}).
The diagonal operator $D_g$ acts independently on the different symmetry
broken ground states of the system and thus commutes with the Hamiltonian;
therefore, we can remove it by choosing the proper gauge for the physical
symmetry according to our definition of phases under symmetries.  If the
symmetry of the initial and the final state overlap on a subsector of the
ground state space, this implies (as for the injective case,
Appendix~\ref{app:stform-inj-sym}) that the 1D representations for this
sector in $D_g$ are the same and thus can be removed by a joint gauge
transformation.  Together, this shows that for the classification of
phases under symmetries in the non-injective case, any symmetry can be
understood as a direct sum over independent sectors, where on each sector
the action of the symmetry is given by 
(\ref{eq:app-noninj-sym:final-form-irrep}), where $h\equiv h(g,\beta)$ is
determined by (\ref{eq:app-noninj-sym:hgamma-from-gbeta}).

\section{\label{appendix:gap}
Robustness of the 2D gap}

Here, we prove the robustness of a gap based on a condition of the form
\begin{equation}
\label{eq:com-gap}
\tilde h_i \tilde h_j+\tilde h_i \tilde h_j
\ge -\tfrac18(1-\Delta_{ij}) (\tilde h_i+\tilde h_j)\ ;
\end{equation}
where we consider a square lattice with $\tilde h_i\ge\openone$ acting on
$2\times 2$ plaquettes, $\Delta_{ij}=\Delta_a$ for directly adjacent
plaquettes $i$, $j$ sharing two spins, and $\Delta_{ij}=\Delta_d$ for
diagonally adjacent plaquettes $i$, $j$ having one spin in common. (In 
Section~\ref{sec:2d-gap}, we have given the simplified version where
$\Delta_a=\Delta_d=\Delta$.) Then, 
\[
\tilde H^2=
\sum_i \underbrace{\tilde h_i^2}_{\ge h_i} + 
\sum_{<ij>} \tilde h_i \tilde h_j +
{\sum}^\prime \underbrace{\tilde h_i \tilde h_j}_{\ge0}
\ge \frac{\Delta_a+\Delta_d}{2} \tilde H
\ ,
\]
which implies a gap in the spectrum of $\tilde H$ between $0$ and
$\Delta=(\Delta_a+\Delta_d)/2>0$, and thus a lower bound $\Delta$ on the
gap of $\tilde H$, cf.~Ref.~\onlinecite{fannes:FCS}.

Let us now study the robustness of \eqref{eq:com-gap} under
$\gamma$-deformation of the Hamiltonian.  Let $h_i$, $h_j$ be projectors
which satisfy $h_ih_j+h_jh_i\ge -\tfrac18(1-\Delta_{ij})(h_i+h_j)$. (The
proof can be modified for the $h_i$ not being projectors.) Let $h_i$ be
supported on systems $AB$, and $h_j$ on systems $BC$, where the number of
sites in systems $A$, $B$, and $C$ is $a$, $b$, and $c=a$,
respectively. (For the square lattice, $a=b=c=2$ for directly neighboring
terms, and $a=c=3$, $b=1$ for diagonally adjacent terms.) With
$Q_\gamma=(1-\gamma)\openone+\gamma Q\le\openone$ as in the
one-dimensional case, let
$\Lambda_{\gamma,X}=\left(Q_\gamma^{-1}\right)^{\otimes x}$, with
$X=A,B,C$ and $x=a,b,c$. Then, the $\gamma$-deformed Hamiltonians are 
\begin{eqnarray*}
h_i(\gamma)&=&(\Lambda_{\gamma,A}\otimes
\Lambda_{\gamma,B})h_i(\Lambda_{\gamma,A}\otimes\Lambda_{\gamma,B})\ ,
\\
h_j(\gamma)&=&(\Lambda_{\gamma,B}\otimes
\Lambda_{\gamma,C})h_j(\Lambda_{\gamma,B}\otimes\Lambda_{\gamma,C})\ .
\end{eqnarray*}
Let us define 
\begin{eqnarray*}
\Theta_\gamma&:=&\Lambda_{\gamma,B}^2-\openone\ge0\ , 
\\
q&:=&\lambda_{\min}(Q)<1\ ,
\\
\mu_\gamma&:=&((1-\gamma)+\gamma \lambda_{\min}(Q))^{-2}\ge1\ ,
\rule[-0.4cm]{0cm}{0.4cm}
\end{eqnarray*}
such that $Q_\gamma^{2a}\ge
\tfrac{1}{\mu_\gamma^a}\openone$, and
$(\mu_\gamma^b-1)\openone\ge\Theta_\gamma$. 
Then, we find  that
\begin{widetext}
        \begin{align*}
        h_i\Lambda_{\gamma,B}^2h_j+ &  h_j\Lambda_{\gamma,B}^2h_i+ 
    \tfrac18
    \mu^a_\gamma \left[ 1-\Delta_{ij}+8(\mu_\gamma^b-1)\right] (h_i\otimes
    \Lambda_{\gamma,C}^{-2}+\Lambda_{\gamma,A}^{-2}\otimes h_j)\\
        &\ge h_i\Lambda_{\gamma,B}^2h_j+h_j\Lambda_{\gamma,B}^2h_i+ 
    \tfrac18
    \left[1-\Delta_{ij}+8(\mu^b_\gamma-1)\right]
    (h_i\otimes
    \openone_C+\openone_A\otimes h_j)\\
    &=
    h_i\Theta_\gamma h_j+h_j\Theta_\gamma h_i+
    h_i\left[ (\mu^b_\gamma-1)\openone\right] h_i+h_j\left[
    (\mu^b_\gamma-1)\openone\right] h_j + 
    h_ih_j+h_jh_i+\tfrac18(1-\Delta_{ij})(h_i+h_j)
    \\
    &\ge
    (h_i+h_j)\Theta_\gamma(h_i+h_j)\ge 0\ .
\end{align*}
\end{widetext}
By multiplying this with $\Lambda_{\gamma,A}\otimes
\Lambda_{\gamma,B}\otimes \Lambda_{\gamma,C}$ from both sides, we obtain 
a lower bound of type \eqref{eq:com-gap} for the $\gamma$-deformed
Hamiltonian,
\begin{equation}
        \label{eq:com-gap-gammadef}
h_i(\gamma) h_j(\gamma)+ h_j(\gamma) h_i(\gamma)\ge
-\tfrac18[1-\Delta_{ij}(\gamma)]
[h_i(\gamma)+h_j(\gamma)]\ ,
\end{equation}
with $\Delta_{ij}(\gamma)=
\mu^a_\gamma\Delta_{ij}+(1+7\mu^a_\gamma-8\mu^{a+b}_\gamma)$.  This can be
used to find an environment of any point in which the system is
still gapped.  In particular, in the case where the isometric parent
Hamiltonian is commuting, and assuming a square lattice, the lower bound
on the spectral gap provided by \eqref{eq:com-gap-gammadef} is 
\[
\Delta(\gamma)=\frac{\Delta_a(\gamma)+\Delta_d(\gamma)}{2}=
    1+4\mu_\gamma^2(1+\mu_\gamma-2\mu_\gamma^2)\ .
\]
This gap vanishes at $\mu_\gamma\approx 1.07$, limiting the maximal deformation of
the isometric tensor to $\lambda_{\min}(Q)/\lambda_{\max}(Q)\approx0.967$.

\end{document}